\documentclass[a4paper,11pt]{iopart}
\usepackage{iopams}
\usepackage{amsfonts} 
\usepackage{epsf}
\usepackage{euscript}
\usepackage{graphicx}
\usepackage{color}
\usepackage{fancyhdr}
\usepackage[indentafter]{titlesec}
\usepackage{etoolbox}
\usepackage{hyperref} 
\usepackage{lipsum}
\usepackage{tocloft}

\usepackage{mathabx}

\begin{document}
 

\title[Asymptotics for the Phase Space Schr\"{o}dinger Equation]{Asymptotics for the Phase Space Schr\"{o}dinger Equation}

\author{Panos D Karageorge$^1$\footnote{\texttt{pkarag}@\texttt{uoc.gr}} and George N Makrakis$^{1,2}$\footnote{\texttt{makrakg}@\texttt{uoc.gr} and \texttt{g.n.makrakis}@\texttt{iacm.forth.gr}}}
\address{$^1$Department of Mathematics and Applied Mathematics, University of Crete, Voutes Campus, 700 13 Heraklion, Greece}
\address{$^2$Institute of Applied and Computational Mathematics, Foundation for Research and Technology, 100 Nikolaou Plastira St, Vassilika Vouton, 700 13 Heraklion, Greece}

\begin{abstract}
We consider semi-classical time evolution for the phase space Schr\"{o}dinger equation. We construct a semi-classical phase space propagator in terms of semi-classical wave packets by the Anisotropic Gaussian Approximation, related to the Nearby Orbit Approximation. We deduce the canonical system in double phase space, related to the Berezin-Shubin-Marinov canonical system and construct a semi-classical asymptotic solution of the Cauchy problem for the phase space Schr\"{o}dinger equation for initial WKB states. We illustrate the method for sub-quadratic potentials in $\mathbb{R}$.
\end{abstract}

\normalsize

\tableofcontents

\section{Introduction}
\label{intro}

\subsection{The General Setting}

Phase space formulations of Quantum Mechanics constitute the sufficient theoretical frame for the description of microscopic physical processes strongly influenced by their external environment, by representing mixed quantum states in terms of phase space quasi-probability distributions, which are used to express expectation values or classical energy densities and fluxes as phase space averages \cite{ZFC}. Despite the relative complexity of these formulations compared to the Schr\"{o}dinger representation, with issues arising such as non-uniqueness of phase space quasi-densities, doubling of variables, non-locality of evolution equations, etc., they prove worth studying in themselves, even outside the context of the theory of Open Quantum Systems, for a series of reasons. 

The phase space is the conceptually natural setting of Quantum Mechanics, more so on its border with Classical Mechanics, the \textit{semi-classical limit,} the range of motions of a given physical system for which the Correspondence Principle is valid \cite{BeSh}. Therein, the correspondence between the theories becomes all the more transparent. Besides the important conceptual reasons for a phase space formulation of Quantum Mechanics, there is a significant methodological reason which dictates such a choice, even for the study of closed physical systems. This is the inherent shortcoming of conventional asymptotic methods to provide global asymptotic solutions for the problem of semi-classical time evolution in a direct fashion\footnote{The complete semi-classical theory of the Cauchy problem for the Schr\"{o}dinger equation, Maslov's theory \cite{Mas1,MaFe,Mas2}, goes around this problem by patching together local semi-classical solutions by means of the semi-classical Fo!
 urier transform; we emphasize that we are considering \textit{direct} methods of obtaining global asymptotic solutions in a given configuration frame-work.}, yielding semi-classical solutions exhibiting singularities at finite times, void of physical content, due to the development of caustics \cite{MaFe}. Caustics are persistent obstacles toward global semi-classical asymptotic solutions of the Schr\"{o}dinger equation in position space and momentum space, alike, while in phase space formulations singularity formation due to caustics is resolved. Another technical reason is that in the context of phase space formulations it is plausible to make a unified approach to the two main classes of semi-classical quantum states, i.e., \textit{WKB} or \textit{Lagrangian states} and \textit{coherent states} (e.g., \cite{HHL}).

The \textit{wave packet representation} is a particular phase space formulation of Quantum Mechanics, first proposed in the works of Torres-Vega and Frederick \cite{ToFr} and thereafter by Harriman \cite{Har}, Chruscinski and Mlodawski \cite{ChMl}, de Gosson \cite{Gos5} and Nazaikinskii \cite{NaSt}, to mention some principal contributions. In its core is the correspondence of pure classical states, i.e., single phase space points, to \textit{semi-classical isotropic Gaussian wave packets}, or \textit{coherent states}, localized at that phase space point on the Heisenberg scale, $O(\hbar^{1/2})$, 
\begin{equation*}
G_{(q,p)}^\hbar(x)=(\pi\hbar)^{-d/4}e^{\frac{i}{\hbar}\Big(\frac{p\cdot q}{2}+p\cdot (x-q)+\frac{i}{2}|x-q|^2\Big)} 
\end{equation*}
where the point $(q,p)\in\mathbb{R}^{2d}$ is called the \textit{base point} of the wave packet.

In the wave packet representation, the configuration space wave function is expressed as a wave packet superposition for the totality of base phase space points
\begin{equation*}\psi(x)=\Big(\frac{1}{2\pi\hbar}\Big)^{d/2}\int\Psi(q,p) \,G_{(q,p)}^\hbar(x)\, dqdp\end{equation*}
where the coefficient of the wave packet superposition is defined as the \textit{phase space wave function}, $\Psi(q,p)$, given, in turn, by
\begin{equation*}\Psi(q,p)=\Big(\frac{1}{2\pi\hbar}\Big)^{d/2}\int \bar G_{(q,p)}^\hbar(x)\psi(x)\, dx \ .\end{equation*}
The above inverse of the wave packet resolution defines the \textit{wave packet transform}, also known as the \textit{Fourier-Bros-Iagolnitzer transform}, closely related to the \textit{Bargmann transform} \cite{Bar1,Bar2,CoFe,NaSt}. 

The dynamics of the phase space wave function satisfies the non-local \textit{phase space Schr\"{o}dinger equation} \cite{Gos1,Gos2,Gos5,Har,ToFr}. As a phase space formulation founded on the wave packet transform, it is simpler than the Wigner-Weyl formulation, insofar as it is linear in the Schr\"{o}dinger representation; it does, however, have its drawbacks, as it cannot account for microscopic systems in significant interaction with their environment, nor does it escape the essential singularity of the semi-classical limit. 

In the wave packet representation, the phase space Schr\"{o}dinger spectral problem is well understood, e.g., through the works of Luef and de Gosson \cite{Gos3}, who have derived the Schr\"{o}dinger spectral equation departing from the spectral equation of Moyal. However, understanding the corresponding \textit{Cauchy problem} for the phase space Schr\"{o}dinger equation, in particular, the problem of semi-classical time evolution remains terra incognita. The deeper understanding of semi-classical time evolution toward a direct theory of semi-classical dynamics in phase space is a challenging and substantial contribution to the theory of phase space formulations of Quantum Mechanics, unifying approaches taken from the theory of semi-classical Fourier integral operators \cite{Dui,Hor,NOSS} to the theory of the Maslov canonical operator \cite{Mas1,MaFe,MaNa,Mas2}. Important contributions in this directions have been made, for example, by Oshmyan et al. and Nazaikinksii et al.!
  \cite{NSS,NaSt}.

The wave packet transform has been implemented in other problems in differential equations besides the Schr\"{o}dinger equation, beyond the frame of Quantum Mechanics, in a variety of settings, where technical issues of diverse nature arise. It has been applied, for example, to problems for the wave equation \cite{GeTa}, while interest grows in its use in the area of Time-Frequency Analysis, in particular, in the construction of Gabor frames \cite{Gro}.

In this article, we focus our attention to the problem of semi-classical time evolution \textit{directly} in phase space, in particular, to semi-classical asymptotic solutions of the Cauchy problem for the \textit{phase space Schr\"{o}dinger equation}, rather than giving phase space representations of semi-classical solutions of the Schr\"{o}dinger equation. 

A fundamental semi-classical approximation of quantum dynamics is the \textit{Gaussian Approximation}, which pertains to approximating quantum evolution of position-momentum localized quantum states with the dynamics of an \textit{individual} Gaussian wave packet translated along a Hamiltonian orbit. This proposition is traced back to the foundational work of Schr\"{o}dinger \cite{Schr}. Semi-Classical Gaussian wave packets provide a natural semi-classical approximation of quantum states, for free motion, as they are localized in phase space on the Heisenberg scale $O(\hbar^{1/2})$, occupying a Planck cell centered at that point, and exhibiting oscillations at the de Broglie wavelength $O(\hbar)$. For a general account on coherent states see \cite{CoRo}. 

The theory of semi-classical wave packet dynamics has profoundly evolved thereafter. Heller \cite{Hel} and Heller et al. \cite{Hub} argued for the use of single isotropic and anisotropic Gaussian wave packets as an approximation to the propagation of initially semi-classical Gaussian wave packets, based on the semi-classical \textit{Nearby Orbit Approximation}, while Huber, Heller and Littlejohn \cite{HHL} showed that isotropic Gaussian wave packet dynamics can stand as a generalization of complex phase WKB semi-classical propagation; an analogous systematic work on isotropic Gaussian wave packet dynamics from the viewpoint of the work of Maslov is that of Bagrov et al. \cite{BBT}, while there are other works along these lines, such as that of Robert \cite{Rob1}, Nazaikinskii et al. \cite{NSS}, Stoyanovsky \cite{Sto}, de Gosson \cite{Gos6}, Faure \cite{Fau1} and Ohsawa et al. \cite{OhLe}. Hagedorn \cite{Hag} showed that an initially Gaussian state retains its Gaussian form w!
 ithin a certain semi-classical timescale, under quantum dynamics. On a different direction, Maslov et al. considered the anisotropic case, generalizing the technique in Quantum Field Theory \cite{MaSh1,MaSh2}. A deep review on the subject of semi-classical wave packet dynamics, from a physical perspective, masterfully touching upon its dynamical and algebraic aspects, is given by Littlejohn \cite{Lit1}.

These ideas have been implemented in the solution of the more general problem of establishing asymptotic solutions of the Schr\"{o}dinger equation or the wave equation, along a given curve, modulated by a Gaussian profile known as \textit{Gaussian beams}. Beginning with the ground-breaking work of Babich and Danilov \cite{BaDa}, who considered asymptotic solutions of the Schr\"{o}dinger equation concentrated along a reference curve, a programme of semi-classical techniques was born. A complete account on Gaussian beams is given by Katchalov et al. \cite{KKL}. Gaussian beams find applications in a wide range of physical problems, in Optics, Acoustics \cite{Kat,QiYi}. 

In the case of anisotropic Gaussian wave packets, we have the following ansatz for semi-classical time evolution \cite{BBT,NSS,Rob1}
\begin{equation*}
G^{Z,\hbar}_{(q,p)}(x,t)=(\pi\hbar)^{-d/4}a(q,p,t)e^{\frac{i}{\hbar}\Big(\frac{p\cdot q}{2}+\mathcal{A}(q,p,t)+p_t\cdot (x-q_t)+\frac{1}{2}(x-q_t)\cdot Z(x-q_t)\Big)}
\end{equation*} 
where $(q_t,p_t)$ is the image of the point $(q,p)$ under the Hamiltonian flow, demanding it to be a semi-classical asymptotic solution of the Cauchy problem, in the following sense \cite{Rob1}
\begin{equation*}
\Big\|\Big(i\hbar\,\frac{\partial}{\partial t}-\widehat H\Big)G^{Z,\hbar}_{(q,p)}(t)\Big\|=O(\hbar^{3/2}) \ , \ \ \hbar\rightarrow 0^+ \ .
\end{equation*}

The quadratic anisotropy form, which satisfies the symmetry and positivity properties $Z^{{\rm T}}=Z$ and ${\rm Im}\, Z\succ 0$, is shown to obey a certain matrix Riccati time evolution equation, equivalent to the dynamics of initially nearby orbits. These dynamics are common to all Gaussian beam asymptotic solutions \cite{KKL,Kel,LiRa,QiYi,Zel1,Zel2}. 

In \cite{Lit1} Littlejohn generalized the works of Heller et al. on the thawed, or anisotropic, Gaussian dynamics for generic initially localized states, by constructing a phase space propagator, as an explicit composition of Weyl shifts and metaplectic operators. Closely related to the work of Heller and Littlejohn, Maslov and Shvedov \cite{MaSh1,MaSh2} suggested the dynamics of anisotropic Gaussian wave packets, giving an alternative representation of the dynamics of the anisotropy quadratic form, as a flow in Siegel upper half-space.

The Gaussian approximation, however, breaks down as localization is lost in an irreversible spreading at a certain semi-classical timescale \cite{Rob1,SVT}, the \textit{Ehrenfest time-scale,} an effect suppressed only for quadratic scalar potentials. Added to the above, the fact that for quadratic potentials the evolution of wave packet \textit{superpositions} of phase space eigenfunctions result in expressions reminiscent of the evolution of single Gaussian wave packets, hints the method of approximating the evolution of a quantum state \textit{by a superposition of semi-classically propagated Gaussian wave packets}. 

We construct a semi-classical phase space propagator based on the \textit{Anisotropic Gaussian Approximation,} which is closely related to the \textit{Nearby Orbit Approximation.} As the Anisotropic Gaussian Approximation is applied for the totality of orbits of the underlying Hamiltonian flow, all of which are taken into account in the wave packet resolution of the phase space wave function, the semi-classical propagator admits generic semi-classical initial data, not just localized ones.

The starting point of this approximation is the resolution of the identity in quantum state space in the over-complete set of coherent states, in particular isotropic Gaussian wave packets, 
\begin{equation*}
{\rm Id}=\Big(\frac{1}{2\pi\hbar}\Big)^{d}\int G_{(q,p)}^\hbar \langle G_{(q,p)}^\hbar,\cdot \rangle \,dqdp
\end{equation*} 
by which we obtain the following representation for the Schr\"{o}dinger flow
\begin{equation*}
U^t=\Big(\frac{1}{2\pi\hbar}\Big)^{d}\int U^t G_{(q,p)}^\hbar \langle G_{(q,p)}^\hbar,\cdot \rangle \,dqdp \ .
\end{equation*}
The approximation itself amounts to an explicit ansatz for the semi-classical propagation of a single wave packet under the Schr\"{o}dinger propagator \cite{NSS,Rob1}, $U^t G_{(q,p)}^\hbar$. The totality of approximations involving single Gaussian dynamics amount to the \textit{Initial Value Representations} of quantum dynamics, closely related to the theory of semi-classical Fourier Integral Operators.

The first work in this direction was that of Herman and Kluk \cite{HeKl}, who argued on the validity of approximating semi-classical evolution by analyzing wave functions by a multitude of non-spreading isotropic Gaussian wave packets, their form held rigid, modulated by some overall amplitude and phase factor, setting off from the van Vleck approximation of the semi-classical propagator. More recently, Rousse and Robert \cite{Rob2,RoSw} assumed a semi-classical time evolution for generic initial data of the Schr\"{o}dinger equation, in terms of a certain semi-classical Fourier integral operator, which is readily identified with the Herman-Kluk propagator, in order to justify this approximation on the basis of estimates for the asymptotic solutions. Other works in the direction of a direct theory of time evolution in the frame of linear representations of the Schr\"{o}dinger equation include the work of Almeida et al. \cite{AlBr}

Besides the traditional field of application of Initial Value Representations in semi-classical schemes, such as Quantum Chemistry, groundbreaking progress in areas such as Quantum Optics (see, e.g., \cite{ZSM}), in atomic optical trapping, bore new interest in Initial Value Representations, and phase space representations in general. In such applications, one is able to generate optical traps or scatterers by multiple LASER pulses, well approximated by linear or parabolic potentials. Dynamics are simplified by additional techniques of LASER cooling, enabling one to focus on the overall orbital motion, by suppressing internal degrees of freedom to their ground quantum states. We also note the work of \cite{CTY} in the field of Theoretical Seismology, where an asymptotic wave group was constructed for the high frequency Cauchy problem for the wave equation, as a model of high frequency acoustic wave propagation in a small depth sub-terrain inhomogeneous medium, by means of th!
 e Isotropic Gaussian Approximation.

The focus of this paper is the semi-classical Cauchy problem for the Weyl-symmetrized phase space Schr\"{o}dinger equation 
\begin{equation*}
i\hbar\,\frac{\partial \Psi}{\partial t}=H\Big(\frac{q}{2}+i\hbar\,\frac{\partial}{\partial p},\frac{p}{2}-i\hbar\,\frac{\partial}{\partial q}\Big)\Psi
\end{equation*}
for uniform, Weyl ordering of the non-commutative operator arguments, for semi-classical initial data in phase space, and the construction of a semi-classical propagator and asymptotic solution based on the \textit{Anisotropic Gaussian Approximation.}

The construction of the semi-classical propagator amounts to analyzing an initial WKB state in the over-complete system, or Weyl-Heisenberg frame, of semi-classical Gaussian wave packets by the wave packet transform and propagating each one, individually, according to the Anisotropic Gaussian Approximation. We consider the evolution as a superposition of propagated wave packets, and consequently express the phase space propagator in terms of propagated and non-propagated wave packets. 

We consider the WKB semi-classical asymptotics of the Weyl symbol of the propagator, as first considered by Berezin and Shubin \cite{BeSh}, in order to determine the corresponding asymptotic form of the phase space propagator kernel. The canonical system derived by Berezin and Shubin, which was determined in the context of asymptotics for the Weyl symbol of the Green operator for the Schr\"{o}dinger equation, comprises of a symmetrized Hamilton-Jacobi equation and a corresponding transport equation, the symmetrized Hamilton-Jacobi equation determined at a later time by Marinov on the basis purely geometric arguments \cite{Mar}.

Based on this canonical system, we derive the canonical system for the phase space propagator kernel, a re-scaled symmetrized Hamilton-Jacobi equation, which we term the \textit{Weyl-symmetrized Hamilton-Jacobi equation,} and a \textit{Weyl-symmetrized transport equation.} We continue, on the basis of the Anisotropic Gaussian Approximation for the kernel of the phase space propagator, to construct the asymptotic solution of the semi-classical Cauchy problem for the phase space Schr\"{o}dinger equation in terms of a semi-classical Fourier integral operator, for WKB initial data. From this superposition form, we re-derive the aforementioned canonical system, the Weyl-symmetrized Hamilton-Jacobi equation and the transport equation for the specific Cauchy problem, whose asymptotic solution is given by the complex WKB method as developed by Maslov \cite{Mas2} in order to yield an expression for the asymptotic solution over the propagated Lagrangian manifold of the initial WKB sta!
 te. We note the connection of the method to the Initial Value Representations, with such schemes as the Herman-Kluk Approximation or the Frozen Gaussian Approximation \cite{HeKl}, as well as the Littlejohn Approximation \cite{Lit2}.

An important result of the analysis of the semi-classical asymptotic analysis is that the Weyl and normal quantization of symbols in double phase space are identified.

Finally, we give detailed illustrations of the method for the short time evolution for scalar sub-quadratic potentials on the real line, pertaining to microscopic physical processes of low energy electron interacting with simple electrostatic potential fields, such as free motion, scattering off a constant electrostatic field and bound motion by a parabolic optical trap.

\subsection{Assumptions and Notational Conventions}

We consider semi-classical time evolution for closed non-relativistic microscopic physical systems comprising of $n$ particles with no spin interactions, such as isolated atomic systems with electrostatic interactions, or closed systems of electron transport in mesoscopic structures under the influence of an electrostatic field.

We assume $\mathbb{R}^d$ as the configuration space with co-ordinates $x=(x_1,\ldots,x_d)$ and $\mathbb{R}^d\oplus\mathbb{R}^d\cong \mathbb{R}^{2d}$ as the phase space, the parallelization of the trivial cotangent bundle of $\mathbb{R}^d$, with canonical co-ordinates $(q,p)=(q_1,\ldots,q_d,p_1,\ldots,p_d)$, or, collectively $X=(q,p)$, while we denote for a different point the canonical co-ordinates $(\eta,\xi)=(\eta_1,\ldots,\eta_d,\xi_1,\ldots,\xi_d)$, or, collectively $Y=(\eta,\xi)$. As a symplectic space, the phase space is equipped with the symplectic form $\omega(X,Y)=X\cdot J Y=q\cdot \xi-p\cdot \eta=\sum_{j=1}^d(q_j \xi_j-p_j\eta_j)$, where $J=\left(\begin{array}{ccc} 0 & I \\ -I & 0 \end{array} \right)$ is the symplectic matrix with respect to the canonical basis. In the dual of the canonical basis we write the symplectic form, or canonical $2$-form, as $\omega=dp\wedge dq=\sum_{j=1}^ddp_j\wedge dq_j$; we denote by $\omega^1=\sum_{j=1}^dp_j \,dq_j$ the normal canonic!
 al $1$-form and by $\omega^1_W=\frac{1}{2}(p\cdot dq-q\cdot dp)=\sum_{j=1}^d\frac{1}{2}(p_j\, dq_j-q_j \,dp_j)$ the Weyl-symmetrized canonical $1$-form \cite{Gos1}. 

We consider smooth autonomous Hamiltonian systems with smooth Hamiltonian function $H$ satisfying the growth condition 
\begin{equation*}\Big|\frac{\partial ^{\alpha+\beta} H}{\partial q^\alpha\partial p^\beta}(q,p)\Big|\leq C_{\alpha\beta}\Big(1+|q|+|p|\Big)^{M_{|\alpha+\beta|}}\end{equation*}
for some constants $C_{\alpha\beta}>0$ and $M_{|\alpha+\beta|}\in\mathbb{R}$, for any pair of multi-indices $\alpha,\beta\in\mathbb{N}^d_0$.

We denote the Hamiltonian flow generated by $H$ by $g^t_H$ and by $(q_t,p_t)=(q_t(q,p),p_t(q,p)):=g^t_H(q,p)$ the terminal point of the orbit with duration $t\geq0$ emanating from the point $(q,p)$ with image $\gamma^t(q,p)=\bigcup_{0\leq \tau\leq t}\{g^\tau(q,p)\}$. As for the dynamical properties of the Hamiltonian, we shall make no assumptions; rather, we shall reversely consider how the validity of the semi-classical approximation is affected by such properties.

Further on notation, for complex entry matrices $A=(a_{kl})$ we write $A^{\rm T}=(a_{lk})$ for its transpose, $\bar A=(\bar a_{kl})$ for its complex conjugate and $A^*=\bar A^{\rm T}$ for its hermitian adjoint. For smooth phase space complex valued functions $f$ we use the notation $\frac{\partial f }{\partial q}$ and $\frac{\partial f }{\partial p}$, for the column vector of partial derivatives, while $\Big(\frac{\partial f }{\partial q}\Big)^{\rm T}$ and $\Big(\frac{\partial f }{\partial p}\Big)^{\rm T}$ for the row vector of partial derivatives, respectively; further, we write $f _{qq}=\Big(\frac{\partial^2f }{\partial q_k\partial q_l}\Big)$, $f _{pp}=\Big(\frac{\partial^2f }{\partial p_k\partial p_l}\Big)$ and $f _{pq}=\Big(\frac{\partial^2f }{\partial q_k\partial p_l}\Big)$, while it is clear that $f _{qp}=f _{pq}^{\rm T}$; for the Hessian matrix we use the block matrix notation $f ''=\left(
\begin{array}{ccc}
f _{qq} & f _{pq} \\
f _{qp} & f _{pp} 
\end{array}
\right)$. 
\\

\begin{table*}
	\centering
		\begin{tabular}{|l|l|} 
		
$X=(q,p),Y=(\eta,\xi),P=(u,v)$ & \scriptsize{phase space points} \\		
$g^t_H$ & \scriptsize{Hamiltonian flow generated by $H$} \\
$(q_t,p_t)$ & \scriptsize{propagation of $(q,p)$ under Hamiltonian flow} \\
$\gamma^t(q,p)$ & \scriptsize{orbit of duration $t$ emanating from $(q,p)$} \\
$\mathfrak{F}$ & \scriptsize{Fock-Bargmann space} \\
$\psi$ & \scriptsize{configuration space wave function} \\ 
$\Psi$ & \scriptsize{phase space wave function} \\
$\mathcal{W}$ & \scriptsize{wave packet transform} \\
$\widehat f$ & \scriptsize{Weyl quantization of physical quantity} $f$\\
$\sigma_W(L)$ & \scriptsize{Weyl symbol of operator} $L$\\
$\widecheck f$ & \scriptsize{wave packet representation of Weyl quantization of physical quantity} $f$ \\
$G_{(q,p)}^\hbar$ & \scriptsize{semi-classical isotropic Gaussian wave packet} \\ 
$Z(t)$ & \scriptsize{anisotropy form} \\
$A(t),B(t)$ & \scriptsize{position and momentum variational forms} \\
$\mathcal{A}(q,p,t)$ & \scriptsize{phase space action}\\
$G_{(q,p)}^{Z,\hbar}(t)$ & \scriptsize{propagated semi-classical anisotropic Gaussian wave packet} \\
$U^t$ & \scriptsize{propagator} \\ 
$K(x,y,t;\hbar)$ & \scriptsize{kernel of} $U^t$\\
$\mathcal{U}^t$ & \scriptsize{phase space propagator} \\ 
$\mathcal{K}(q,p,\eta,\xi,t;\hbar)$ & \scriptsize{kernel of} $\mathcal{U}^t$\\
$U_{sc}^t$ & \scriptsize{semi-classical approximation of $U^t$} \\ 
$K_{sc}(x,y,t;\hbar)$ & \scriptsize{kernel of} $U^t_{sc}$\\
$\mathcal{U}_{sc}^t$ & \scriptsize{semi-classical phase space propagator} \\ 
$\mathcal{K}_{sc}(q,p,\eta,\xi,t;\hbar)$ & \scriptsize{kernel of} $\mathcal{U}^t_{sc}$\\
$Q(t)$ & \scriptsize{double phase space anisotropy form} \\
$C(t),D(t)$ & \scriptsize{double phase space position and momentum variational forms} \\
$\psi^\hbar(t)$ & \scriptsize{asymptotic solution of Schr\"{o}dinger equation} \\ 
$\Psi^\hbar(t)$ & \scriptsize{asymptotic solution of phase space Schr\"{o}dinger equation} \\
\end{tabular}	
\caption{Reference table of basic symbol notations.}
\label{tab:sksksksk}
\end{table*}

\section{The Phase Space Schr\"{o}dinger Equation}

\subsection{The Wave Packet Transform}
The \textit{wave packet representation} is a linear representation of Quantum Mechanics over phase space, first proposed by Torres-Vega and Frederick \cite{ToFr}, elaborated by other authors subsequently, e.g., Harriman \cite{Har} and de Gosson \cite{Gos5}. It is related to the Schr\"{o}dinger representation by means of a linear unitary operator, the \textit{wave packet transform} \cite{NSS,NaSt,ToFr}, which maps position space wave functions, $\psi(x)$, to phase space wave functions, $\Psi(q,p)$,
\begin{equation}
\mathcal{W}:L^2(\mathbb{R}^d,\mathbb{C};dx)\rightarrow L^2(\mathbb{R}^{2d},\mathbb{C};dqdp)\,|\,\psi\mapsto\Psi=\mathcal{W}\psi \ . 
\end{equation}
It is defined explicitly by the following integral transform \cite{NSS,NaSt,ToFr}
\begin{equation}
\Psi(q,p)=\mathcal{W}\psi(q,p)=\Big(\frac{1}{2\pi\hbar}\Big)^{d/2}\int\bar G_{(q,p)}^\hbar(x)\psi(x)\,dx 
\label{eq:wpt}
\end{equation}
its kernel being the complex conjugated semi-classical isotropic Gaussian wave packet with base point $(q,p)\in\mathbb{R}^{2d}$
\begin{equation}
G_{(q,p)}^\hbar(x)=(\pi\hbar)^{-d/4}\exp\frac{i}{\hbar}\Big(\frac{p\cdot q}{2}+p\cdot (x-q)+\frac{i}{2}|x-q|^2\Big) \ .
\label{eq:g}
\end{equation}

As a linear operator between Hilbert spaces, the wave packet transform is not a bijection; its image is a sub-space 
\begin{equation}
\mathfrak{F}\subset L^2(\mathbb{R}^{2d})
\end{equation} 
the \textit{Fock-Bargmann space}, defined by the \textit{Fock-Bargmann constraint} \cite{ChMl,Gos2,Har,NSS,NaSt,ToFr}
\begin{equation}
\Psi\in\mathfrak{F}\,\iff\,\Bigg(\Big(\frac{q}{2}-i\hbar\,\frac{\partial}{\partial p}\Big)-i\Big(\frac{p}{2}+i\hbar\,\frac{\partial}{\partial q}\Big)\Bigg)\Psi=0
\label{eq:FBc}
\end{equation}
equivalent to the Cauchy-Riemann relations
\begin{equation}
\Big(\frac{\partial}{\partial q}-i\frac{\partial}{\partial p} \Big)\Big(e^{\frac{1}{2\hbar}(ip\cdot q-|p|^2)}\Psi\Big)=0
\end{equation}
implying that \textit{not all square integrable phase space functions constitute phase space wave functions} \cite{Bar1,Gos2,NSS,NaSt,ToFr}; only Gaussian-weighted square integrable analytic functions in the variable $q-ip\in\mathbb{C}^d$.

\subsection{Semi-Classical Weyl Operators in Phase Space}

For physical quantities of appropriate smoothness and growth, we define their quantization\footnote{We choose the\textit{ Weyl quantization,} or \textit{Weyl ordering.}}, $\widehat f={\rm Op}_W(f)$, as
\begin{equation}
\widehat f=f\Big(\stackrel{\boldsymbol{\omega}}{x},-i\hbar\,\stackrel{\boldsymbol{\omega}}{\frac{\partial}{\partial x}}\Big):=f\Big(\frac{\stackrel{1}{x}+\stackrel{3}{x}}{2},-i\hbar\,\stackrel{2}{\frac{\partial}{\partial x}}\Big)
\end{equation}
the Feynman indexes stating the order of action of the symbol non-commuting operator arguments \cite{Fey,KaMa,Mas1,MaNa}. The operator $\widehat f$ is a \textit{semi-classical Weyl operator} defined by the integral representation 
\begin{equation}
\widehat f\psi(x)=\Big(\frac{1}{2\pi\hbar}\Big)^d\int e^{\frac{i}{\hbar}p\cdot (x-q)}f\Big(\frac{x+q}{2},p\Big)\psi(q)\,dqdp
\label{eq:irweyl}
\end{equation}
with symbol, in phase space, $\sigma_W(\widehat f)=f$.

In order to pose the Schr\"{o}dinger equation and the problem of time evolution in phase space, i.e., in the wave packet representation, one must define the quantization of a physical quantity in the wave packet representation
\begin{equation}
\widecheck f:=\mathcal{W}\widehat f \mathcal{W}^{-1} \ .
\end{equation}
 One may use the above in order to construct the phase space Schr\"{o}dinger equation by conjugating with the wave packet transform 
\begin{equation}
i\hbar\,\frac{\partial \psi}{\partial t}=\widehat H\psi=0 \ \iff \ i\hbar\,\frac{\partial \Psi}{\partial t}=\widecheck H\Psi
\end{equation}
where $\Psi=\mathcal{W}\psi$ and $\widecheck H=\mathcal{W}\widehat H \mathcal{W}^{-1}$. 

In general, we define $\widecheck f$ as a semi-classical Weyl operator in phase space according to the phase space Weyl calculus developed by de Gosson \cite{Gos1,Gos2}, meaning that it acts on appropriate phase space wave functions, with a double phase space symbol. In particular, we have 
\begin{equation}
\widecheck f:=\mathcal{W}f\Big(\stackrel{\boldsymbol{\omega}}{x},-i\hbar\,\stackrel{\boldsymbol{\omega}}{\frac{\partial}{\partial x}}\Big)\mathcal{W}^{-1}=f\Big(\stackrel{\boldsymbol{\omega}}{\frac{q}{2}}+i\hbar\,\stackrel{\boldsymbol{\omega}}{\frac{\partial}{\partial p}},\stackrel{\boldsymbol{\omega}}{\frac{p}{2}}-i\hbar\,\stackrel{\boldsymbol{\omega}}{\frac{\partial}{\partial q}}\Big)
\end{equation}
which is a semi-classical Weyl operator with symbol in the double phase space
\begin{equation}
\sigma_W(\widecheck f)(q,p,u,v)=f\Big(\frac{q}{2}-v,\frac{p}{2}+u\Big)
\end{equation}
where $(q,p)=X\in\mathbb{R}^{2d}$ and $(u,v)=P\in\mathbb{R}^{2d}$, meaning that 
\begin{equation}
f\Big(\stackrel{\boldsymbol{\omega}}{\frac{q}{2}}+i\hbar\,\stackrel{\boldsymbol{\omega}}{\frac{\partial}{\partial p}},\stackrel{\boldsymbol{\omega}}{\frac{p}{2}}-i\hbar\,\stackrel{\boldsymbol{\omega}}{\frac{\partial}{\partial q}}\Big)=\sigma_W(\widecheck f)\Big(\stackrel{\boldsymbol{\omega}}{q},\stackrel{\boldsymbol{\omega}}{p},-i\hbar\, \stackrel{\boldsymbol{\omega}}{\frac{\partial}{\partial q}},-i\hbar\, \stackrel{\boldsymbol{\omega}}{\frac{\partial}{\partial p}}\Big) \ .
\end{equation}
The corresponding integral representation, analogous to (\ref{eq:irweyl}), is
\begin{equation}
\widecheck f\Psi(X)=\Big(\frac{1}{2\pi\hbar}\Big)^{2d}\int e^{\frac{i}{\hbar}P\cdot (X-Y)}\sigma_W(\widecheck f)\Big(\frac{X+Y}{2},P\Big)\Psi(Y)\, dYdP \ .
\end{equation}

One may think of the wave packet representation of physical quantities in the Weyl quantization equivalent to a direct phase space quantization by means of the formal substitution 
\begin{equation}
q\rightarrow \frac{q}{2}+i\hbar\,\frac{\partial}{\partial p} \ , \ \ p\rightarrow \frac{p}{2}-i\hbar\,\frac{\partial}{\partial q}
\end{equation}
the \textit{Bopp shifts} \cite{Bop}, while maintaining a uniform ordering of non-commutative operators.

\subsection{The Cauchy Problem for the Phase Space Schr\"{o}dinger Equation}

We begin by considering the problem of semi-classical time evolution, or the semi-classical Cauchy problem for the Schr\"{o}dinger equation in the Weyl quantization (see \ref{QuRep}) for semi-classical initial data $\psi_0$.

The problem reads
\begin{equation}
\Big(i\hbar\,\frac{\partial}{\partial t}-\widehat H\Big)\psi(t)=0 \ , \ \ t\in [0,T]
\label{eq:ivp}
\end{equation}
with initial condition
\begin{equation}
\psi(0)=\psi_0 \ .
\end{equation}

We shall comment on the constraints imposed by the dynamical properties of the Hamiltonian flow on the admissible duration $T$ for the semi-classical approximation we assume.

The wave packet representation of the Cauchy problem for the \textit{phase space Schr\"{o}dinger equation} reads
\begin{equation}
\Big(i\hbar\,\frac{\partial}{\partial t}-\widecheck H\Big)\Psi(t)=0 \ , \ \ t\in [0,T] \label{eq:pivp}
\end{equation}
with initial condition
\begin{equation}
\Psi(0)=\Psi _0
\end{equation}
for generic initial data $\Psi _0\in\mathfrak{F}$, for the phase space wave function $\Psi=\mathcal{W}\psi$. 

For the semi-classical time evolution problem, the wave packet transform retains the essential singularity of the equation itself, as well as of the initial data, as will be shown in section \ref{solutions} for WKB states as prototype semi-classical states. This renders the phase space image of the initial problem semi-classically singular as well.

A notable difference from the Schr\"{o}dinger equation, in the case of the standard form Hamiltonian, $H(q,p)=|p|^2+V(q)$, arises from the potential term; for non-polynomial potentials, this term introduces non-locality as well as transport effects in phase space dynamics, both common features in phase space evolution equations \cite{ChMl}, such as the von Neumann equation \cite{ZFC}. 

The solution of the Cauchy problem for the Schr\"{o}dinger equation ($\ref{eq:ivp}$), with initial data $\psi_0 \in L^2(\mathbb{R}^{d})$, is given by the action of the Schr\"{o}dinger flow \cite{BeSh,FeHi}
\begin{equation}
\psi (x,t)= U^t\psi _0(x)=\int K(x,y,t;\hbar)\psi _0(y)\,dy 
\label{eq:Kprop}
\end{equation}
$K$ being the \textit{Schr\"{o}dinger propagator}, the kernel of $U^t$.

The Schr\"{o}dinger flow itself evolves according to the dynamics \cite{BeSh} 
\begin{equation}
\Big(i\hbar\,\frac{d}{dt}-\widehat H \Big)U^t=0 \ , \ \ t\in [0,T]
\label{eq:eqprop}
\end{equation}
with initial condition
\begin{equation}
U^0={\rm Id}
\end{equation}
comprising a unitary group\footnote{This is true as the Hamiltonian flow is autonomous.} \cite{BeSh} 
\begin{equation}
\{U^t\}_{t\in\mathbb{R}}=\{e^{-\frac{i}{\hbar}t\widehat H }\}_{t\in\mathbb{R}} 
\label{eq:Ut}
\end{equation}
in the sense that it satisfies the group composition property, for $t,s\in\mathbb{R}$,
\begin{equation}
U^t U^{s}=U^{t+s}
\end{equation}
and the unitarity property, for $t\in\mathbb{R}$, 
\begin{equation}
(U^t)^*=(U^t)^{-1}= U^{-t}\ .
\end{equation}

The kernel of the propagator, $K$, is a fundamental solution of the Schr\"{o}dinger equation \cite{BeSh} in the sense that 
\begin{equation}
\Big(i\hbar\,\frac{\partial}{\partial t}-\widehat H \Big)K(x,y,t;\hbar)=i\hbar\,\delta(t)\delta(x-y) 
\label{eq:fund}
\end{equation}
where the operator acts on the first argument, with initial condition
\begin{equation}
K(x,y,0;\hbar)=\delta(x-y) \ . 
\end{equation}

By beginning with the completeness relation of the semi-classical isotropic Gaussian wave packets, a resolution of the identity in $L^2(\mathbb{R}^d)$ \cite{Rob1}, 
\begin{equation}
\Big(\frac{1}{2\pi\hbar}\Big)^{d}\int\bar G_{(q,p)}^\hbar(x) G_{(q,p)}^\hbar(y)\,dqdp=\delta(x-y)  
\label{eq:completeness}
\end{equation}
we acquire a phase space resolution of the kernel of the propagator
\begin{equation}
K(x,y,t;\hbar)=\Big(\frac{1}{2\pi\hbar}\Big)^{d}\int U^tG_{(q,p)}^\hbar(x)\bar G_{(q,p)}^\hbar(y)\,dqdp \ .
\end{equation}

By applying the wave packet transform ($\ref{eq:wpt}$) on the representation formula ($\ref{eq:Kprop}$), we derive the phase space propagator, $\mathcal{U}^t$, which gives the solution of the Cauchy problem for the phase space Schr\"{o}dinger equation with initial data $\Psi_0 \in\mathfrak{F}$, by the action 
\begin{equation}
\Psi (q,p,t)=\mathcal{U}^t\Psi _0(q,p)=\int \mathcal{K}(q,p,\eta,\xi,t;\hbar)\Psi _0(\eta,\xi)\,d\eta d\xi \ .
\label{eq:pswf}
\end{equation}
The kernel $\mathcal{K}$ has the following representation 
\begin{eqnarray}
\mathcal{K}(q,p,\eta,\xi,t;\hbar)=\Big(\frac{1}{2\pi\hbar}\Big)^{d} \int\!\!\int \bar G_{(q,p)} (x)G_{(\eta,\xi)} (y)K(x,y,t;\hbar)\,dx dy
\label{eq:psprop}
\end{eqnarray}
while the inverse relation is 
\begin{eqnarray}
\fl K(x,y,t;\hbar)=\Big(\frac{1}{2\pi\hbar}\Big)^{d} \int\!\!\int \bar G_{(q,p)} (x)G_{(\eta,\xi)} (y)\mathcal{K}(q,p,\eta,\xi,t;\hbar)\,dqdpd\eta d\xi  \ .
\end{eqnarray}
It can be easily checked that ($\ref{eq:pswf}$) solves the problem ($\ref{eq:pivp}$) for generic initial data $\Psi_0\in \mathfrak{F}$.

Time evolution operator in phase space evolves according to unitary dynamics as well, induced by the self-adjoint Hamiltonian function $\widecheck H$.

We consider the Cauchy problem for the Weyl-symmetrized phase space Schr\"{o}dinger equation (see \ref{QuRep})
\begin{equation}
\Big(i\hbar\,\frac{\partial}{\partial t}-\widecheck H\Big)\Psi(t)=0 \ , \ \ t\in [0,T] \label{eq:pivp}
\end{equation}
with initial condition
\begin{equation}
\Psi(0)=\Psi _0
\end{equation}
with $\Psi _0=\mathcal{W}\psi _0$, where $\psi_0$ is a semi-classical state, for the phase space wave function $\Psi=\mathcal{W}\psi$. For the semi-classical time evolution problem, the wave packet transform retains the essential singularity of the equation itself, as well as of the initial data.

The time evolution operator in phase space evolves according to unitary dynamics 
\begin{equation}
\Big(i\hbar\,\frac{d}{dt}-\widecheck H \Big)\mathcal{U}^t=0 \ , \ \ t\in [0,T]
\end{equation}
with initial condition
\begin{equation}
\mathcal{U}^0={\rm Id}
\end{equation}
comprising a unitary group 
\begin{equation}
\{\mathcal{U}^t\}_{t\in\mathbb{R}}=\{e^{-\frac{i}{\hbar}t\widecheck H }\}_{t\in\mathbb{R}} 
\label{eq:Ut}
\end{equation}
in the sense that it satisfies the group composition property, for $t,s\in\mathbb{R}$,
\begin{equation}
\mathcal{U}^t\mathcal{U}^{s}=\mathcal{U}^{t+s}
\label{eq:U}
\end{equation}
and the unitarity property, for $t\in\mathbb{R}$, 
\begin{equation}
(\mathcal{U}^t)^*=(\mathcal{U}^t)^{-1}= \mathcal{U}^{-t} \ .
\end{equation}

Conjugating (\ref{eq:fund}) by the wave packet transform, we get the analogous equation for the kernel of the phase space propagator
\begin{equation}
\Big(i\hbar\,\frac{\partial}{\partial t}-\widecheck H \Big)\mathcal{K}(q,p,\eta,\xi,t;\hbar)=i\hbar\,\delta(t)\,\langle G_{(q,p)}^\hbar,G_{(\eta,\xi)}^\hbar\rangle
\end{equation}
where the operator acts on the first arguments, with initial condition
\begin{equation} 
\fl \mathcal{K}(q,p,\eta,\xi,0;\hbar)=\Big(\frac{1}{2\pi\hbar}\Big)^{d} \int \bar G_{(q,p)}^\hbar(x)G_{(\eta,\xi)}^\hbar(x)\,dx =:b(q,p,\eta,\xi;\hbar) 
\end{equation}
the so-called \textit{Bergmann kernel} \cite{Bar1,Bar2,NSS}. Due to the Gaussian integration in ($\ref{eq:psprop}$), the kernel $\mathcal{K}$ has stronger smoothness properties than $K$ across the hyper-plane $(q,p)=(\eta,\xi)$, while for $t\rightarrow 0^+$ it does not converge weakly to a Dirac distribution, but rather, it is a Dirac mollifier on the Heisenberg scale; the Bergmann kernel possess the reproducing property, due to an underlying analytic structure called the Bergmann structure of $\mathfrak{F}$,
\begin{equation} 
\int b(q,p,\eta,\xi;\hbar) \Psi(\eta,\xi)\,d\eta d\xi = \Psi(q,p)
\end{equation}
for any $\Psi\in\mathfrak{F}$ (see \ref{WPTr}).

\section{The Anisotropic Gaussian Approximation}

We construct a semi-classical asymptotic propagator on the basis of the \textit{Anisotropic Gaussian Approximation}, a semi-classical approximation of wave packet dynamics. To this end, we consider the propagation of semi-classical Gaussian wave packets in configuration space, in particular, the \textit{semi-classical anisotropic Gaussian wave packet}, following, basically, the work of Robert \cite{Rob1}, Nazaikinskii et al. \cite{NSS}, Belov et al. \cite{BBT}, based on the independent contributions of others (see section \ref{intro}). 

As we shall see in what follows, the semi-classical dynamics of anisotropic Gaussian wave packets in configuration space relies on the \textit{variational system} of the Hamiltonian flow, which is the basis for the \textit{Nearby Orbit Approximation} (see, e.g., \cite{Sto}).

\subsection{Semi-Classical Wave Packet Dynamics}
 
The attempt for semi-classical description of the dynamics that has been termed as \textit{Initial Value Representations} \cite{HeKl} involve superpositions of initial semi-classical states in wave packets, $G_{(q,p)}^\hbar(x)$, propagating them, in a certain approximation along the Hamiltonian orbit $\gamma^t(q,p)$ emanating from the point $(q,p)$ and subsequently superposing the evolved wave packets with respect to the totality of base points, i.e., initial phase space points $(q,p)\in\mathbb{R}^{2d}$.

We consider the \textit{Anisotropic Gaussian Approximation} in order to construct a semi-classical asymptotic phase space propagator, considering linear wave packet superpositions, using as a `building block' the semi-classical propagation of a single isotropic Gaussian wave packet, $U^tG_{(q,p)}^\hbar$, which retains its Gaussian wave packet form, yet acquires an anisotropy in its phase. In particular, in this approximation, the initial state
\begin{equation}
G_{(q,p)}^\hbar(x) =(\pi\hbar)^{-d/4}\exp\frac{i}{\hbar}\Big(\frac{p\cdot q}{2}+p\cdot (x-q)+\frac{i}{2}|x-q|^2\Big) 
\end{equation}
is evolved semi-classically to the anisotropic Gaussian wave packet moving along the trajectory $x=q_t$, and semi-classically concentrated on that point, on the Heisenberg scale \cite{BaDa,NSS,Sto}. 

We begin by considering the following ansatz (see, e.g., \cite{NSS,Rob1})
\begin{equation}
\fl G^{Z,\hbar}_{(q,p)}(x,t)=(\pi\hbar)^{-d/4}a(q,p,t) e^{\frac{i}{\hbar}\Big(\frac{p\cdot q}{2}+\mathcal{A}(q,p,t)+p_t\cdot (x-q_t)+\frac{1}{2}(x-q_t)\cdot Z(q,p,t)(x-q_t)\Big)} 
\label{eq:MSwp}
\end{equation}
demanding it to be an asymptotic solution of the Cauchy problem in the sense that 
\begin{equation}
\Big\|\Big(i\hbar\,\frac{\partial}{\partial t}-\widehat H \Big)G^{Z,\hbar}_{(q,p)}(t)\Big\|=O(\hbar^{3/2}) \ , \ \ \hbar\rightarrow 0^+
\end{equation}
which induces particular dynamics for the `parametrizing' functions, constituting the \textit{characteristic system;} $(q_t,p_t)$ are considered along a simple smooth curve in phase space, $\mathcal{A}(q,p,t)$ is a real valued function, while the matrix $Z={\rm Re}\,Z+i\,{\rm Im}\,Z$ is symmetric 
\begin{equation}
Z^{{\rm T}}=Z
\end{equation}
and has positive definite imaginary part
\begin{equation}
{\rm Im}\,Z\succ 0 \ .
\end{equation}


By substituting the ansatz ($\ref{eq:MSwp}$) in the Schr\"{o}dinger equation, and demanding that it be an asymptotic solution in the above sense, we arrive at a system of differential equations along the curve defined by $\gamma(t)=(q_t,p_t)$ \cite{Lit1,NSS,Rob1,Sto}, the \textit{characteristic system} 
\begin{eqnarray} 
\fl O(1) & \ \ \ \ \ \ \ \  \frac{d\mathcal{A}}{dt}=p_t\cdot \frac{dq_t}{dt}-H \ , & \ \ \ \ \ \ \ \ \mathcal{A}(0)=0 \\ \nonumber
\fl O(\hbar^{1/2}) & \ \ \ \ \ \ \ \  \frac{dq_t}{dt}=\frac{\partial H}{\partial p} \ , \ \ \frac{dp_t}{dt}=-\frac{\partial H}{\partial q} \ ,  & \ \ \ \ \ \ \ \ (q_0,p_0)=(q,p)\\ \nonumber
\fl O(\hbar)& \ \ \ \ \ \ \ \ \frac{da}{dt}+\frac{1}{2}{\rm tr}\Big(H_{pp}Z+H_{pq}\Big)a=0 \ ,  &\ \ \ \ \ \ \ \ a(0)=1\\ \nonumber
\fl & \ \ \ \ \ \ \ \ \frac{dZ}{dt}+ZH_{pp}Z+H_{qp}Z+ZH_{pq}+H_{qq}=0 \ ,  & \ \ \ \ \ \ \ \ Z(0)=iI \ . 
\end{eqnarray}

It is clear that the curve $\gamma(t)=(q_t,p_t)$, along which the above system is defined, the \textit{bi-characteristic} of the system, is the Hamiltonian orbit $\gamma^t(q,p)$, as can be seen explicitly by the by the $O(\hbar^{1/2})$ equations, the equations of Hamilton.

The phase $\mathcal{A}$ is the \textit{phase space action,} closely related to the characteristic function of Hamilton
\begin{equation}
\mathcal{A}(q,p,t)=\int_{\gamma^t(q,p)}\omega^1-H(q,p)\,t=\int\displaylimits_0^tp_\tau\cdot \frac{d q_\tau}{d\tau} \,d\tau-H(q,p)\,t \ .
\label{eq:action}
\end{equation}
The following properties of the phase space action shall be of use in what follows,
\begin{equation}
\frac{\partial \mathcal{A}}{\partial q}=-p+p_t^{{\rm T}}\,\frac{\partial q_t}{\partial q} \ , \ \ \frac{\partial \mathcal{A}}{\partial p}=p_t^{{\rm T}}\,\frac{\partial q_t}{\partial p} \ .
\label{eq:dA}
\end{equation} 

The \textit{amplitude} $a$, satisfying the second order in $\hbar^{1/2}$ transport equation, is given by
\begin{equation}
a(q,p,t)=\exp\Bigg(-\frac{1}{2}\int\displaylimits_0^t{\rm tr}\Big(H_{pp}Z+H_{pq}\Big)\,d\tau\Bigg)
\end{equation}
where the Hessian elements are taken along the transported point $(q_t,p_t)$. 

The quadratic form, $Z$, satisfying the second order in $\hbar^{1/2}$ matrix Riccati equation, the \textit{anisotropy form} \cite{Dar,HHL,NSS,Rob1,Sto}, essentially controls the direction, the shape and spreading of the propagated state.

\subsection{Nearby Orbit Approximation: Variational System and Matrix Riccati Flow}

The wave packet form of $G^{Z,\hbar}_{(q,p)}$ is guaranteed, as by beginning at $Z(0)=iI$, $Z$ remains symmetric with positive definite imaginary part for all times. We stress that the Riccati flow guarantees unitarity of the semi-classical Schr\"{o}dinger flow, in the sense that $\|G^{Z,\hbar}_{(q,p)}(t)\|=1$ for all times, $t\geq 0$. 

As, by assumption, $G^{Z,\hbar}_{(q,p)}\in L^2(\mathbb{R}^d)$, we have $Z(t)^{{\rm T}}=Z(t)$ and ${\rm Im}\,Z(t)\succ 0$, the Riccati equation must define a flow in the \textit{Siegel upper half-space},
\begin{equation}
\Sigma _d:=\{Z\in\mathbb{C}^{d\times d}|\, Z^{{\rm T}}=Z \ , \ \ {\rm Im}\,Z\succ 0\} 
\end{equation}
which is a real $d(d+1)$-dimensional symmetric space, an analogue of the complex upper half-plane $\Sigma _1=\{z\in\mathbb{C}|\,{\rm Im}\,z>0\}$ \cite{OhLe,Sto}.

As noted by Huber et al. \cite{HHL}, in anisotropic Gaussian wave packet dynamics the spatio-temporal variation of its localization is dictated by certain matrix Riccati dynamics, which in turn are related to the separation dynamics of initially nearby Hamiltonian orbits. These dynamics are essentially common to all Gaussian narrow beam dynamics, i.e., dynamics for asymptotic solutions concentrated with a Gaussian profile about the characteristics \cite{Kat}.

The Riccati equation for the anisotropy form is equivalent to the \textit{variational system} governing the stability of the Hamiltonian flow \cite{HHL,NSS,Rob1,Sto}. 

By varying the equations of Hamilton
\begin{equation}
\frac{dq_t}{dt}=\frac{\partial H}{\partial p} \ , \ \ \frac{dp_t}{dt}=-\frac{\partial H}{\partial q} 
\end{equation}
with repsect to the initial points $(q,p)$ and taking appropriate complex combinations, we obtain the variational system, which, following \cite{NSS}, 
\begin{equation}
\frac{d}{dt}\left(
\begin{array}{ccc}
A \\
B 
\end{array}\right)=\left(\begin{array}{ccc}
H_{pq} & H_{pp} \\
-H_{qq} & -H_{qp} 
\end{array}\right)\left(
\begin{array}{ccc}
A \\
B 
\end{array}\right) 
\label{eq:varsys}
\end{equation}
where the Hessian elements evaluated along the orbit, at $(q_t,p_t)$. 

Following Maslov \cite{Mas2}, we coin the term \textit{variational forms} for the solutions of the variational system, in particular, \textit{position variational form} for $A$ and \textit{momentum variational form} for $B$. The anisotropy form decomposes with respect to the variational forms as 
\begin{equation}
Z=BA^{-1}
\end{equation}
where $Z=Z(q,p,t)$ or $Z(t)$, and $A=A(q,p,t)$ or $A(t)$ and $B=B(q,p,t)$ or $B(t)$ respectively. The initial condition $G_{(q,p)}^{Z(0),\hbar}(x,0)=G_{(q,p)}^\hbar(x)$ determines the initial condition for the anisotropy form, $Z(0)=iI$, which yields $B(0)A(0)^{-1}=iI$, so that there is a left multiplication by the unitary group arbitrariness in the definition of the variational forms; we may take the following initial conditions for the variational system
\begin{equation}
A(0)=U \ , \ \ B(0)=iU
\end{equation}
where $U\in{\rm U}(d)$.

The variational system has a unique global solution, bounded away from zero \cite{NaSt}. A specific solution of the variational system, for the choice $\mathfrak{A}(0)=I$ and $\mathfrak{B}(0)=iI$, which we term Hamiltonian gauge, is 
\begin{equation}
\mathfrak{A}(t)=A(t)U^{-1}=\frac{\partial q_t}{\partial q}+i\frac{\partial q_t}{\partial p} \ , \ \ \mathfrak{B}(t)=B(t)U^{-1}=\frac{\partial p_t}{\partial q}+i\frac{\partial p_t}{\partial p}
\end{equation}
so that 
\begin{equation}
Z=BA^{-1}=\mathfrak{B}\mathfrak{A}^{-1} \ .
\end{equation}
Thus, the anisotropy form may be represented as 
\begin{equation}
Z(t)=\Big(\frac{\partial p_t}{\partial q}+i\frac{\partial p_t}{\partial p}\Big)\Big(\frac{\partial q_t}{\partial q}+i\frac{\partial q_t}{\partial p}\Big)^{-1} \ .
\label{eq:Zdyn}
\end{equation}

\subsection{Variational Form Identities}
\label{Relations}

The development of the algebraic relations between variational forms which we present is of methodological importance in asymptotic calculations (see, e.g., \cite{Hag}). The underlying structure of the dynamical constraints between the position and momentum variational forms is the symplectic invariance 
\begin{equation}
\Big(\frac{\partial X_t}{\partial X}\Big)^{{\rm T}}J\,\frac{\partial X_t}{\partial X}=J
\end{equation}
where $X=(q,p)$ and $X_t=g^t_HX=g^t_H(q,p)$, which actually produces the canonical Lagrangian relations along the flow.

The primary algebraic constraints on the anisotropy form are \textit{symmetry}, $Z^{\rm T}=Z$, and \textit{positivity of the imaginary part}, ${\rm Im}\,Z\succ 0$. These are equivalently expressed in terms of the variational forms as 
\begin{equation}
A^{\rm T}B-B^{\rm T}A=0
\end{equation}
and
\begin{equation}
{\rm Im}\,BA^{-1}\succ 0 \ .
\end{equation}

As mentioned, the symplectic character of the Hamiltonian flow bares dynamical relations on the variational forms, by noting that they are expressed modulo right multiplication by a non-zero unitary matrix in an explicit dynamical form, which we term the Hamiltonian gauge
\begin{equation}
\mathfrak{A}(t)=A(t)U^{-1}=\frac{\partial q_t}{\partial q}+i\frac{\partial q_t}{\partial p} \ , \ \ \mathfrak{B}(t)=B(t)U^{-1}=\frac{\partial p_t}{\partial q}+i\frac{\partial p_t}{\partial p} \ .
\end{equation}

These relations are derived by expressing the Poisson and Lagrange canonical relations for the canonical co-ordinates $(q_t,p_t)$ and $(q,p)$, respectively, 
\begin{equation}
\{q_{t\,k},p_{t\,l}\}=\delta_{kl} \ , \ \ \{q_{t \, k},q_{t \, l}\}=0 \ , \ \ \{p_{t \, k},p_{t \, l}\}=0 \ , \ \ k,l=1,\ldots,d
\end{equation}
and 
\begin{equation}
[\![q_k,p_l]\!]=\delta_{kl} \ , \ \ [\![q_k,q_l]\!]=0 \ , \ \ [\![p_k,p_l]\!]=0 \ , \ \ k,l=1,\ldots,d
\end{equation}
where the Poisson and Lagrange brackets of smooth functions $f_1,\ldots,f_d,g_1,\ldots,g_d$ of the canonical co-ordinates $(q,p)$ are \cite{Arn}
\begin{equation}
\{f_k,g_l\}:=\sum_{j=1}^d\Big(\frac{\partial f_{k}}{\partial q_j}\frac{\partial g_{l}}{\partial p_j}-\frac{\partial f_{k}}{\partial p_j}\frac{\partial g_{l}}{\partial q_j}\Big) \ , \ \ k,l=1,\ldots,d
\end{equation}
and the Lagrange brackets of the image canonical co-ordinates $(\eta,\xi)$ with respect to the canonical co-ordinates $(q_t,p_t)$ are \cite{Arn} 
\begin{equation}
[\![\eta_k,\xi_l]\!]:=\sum_{j=1}^d\Big(\frac{\partial q_{t\,j}}{\partial \eta_k}\frac{\partial p_{t\,j}}{\partial \xi_l}-\frac{\partial q_{t\,j}}{\partial \xi_l}\frac{\partial p_{t\,j}}{\partial \eta_k}\Big) \ , \ \ k,l=1,\ldots,d \ .
\end{equation}

The expression of the dynamical coefficients in terms of the variational forms is 
\begin{equation}
\frac{\partial q_t}{\partial q}=\frac{\mathfrak{A}+\bar \mathfrak{A}}{2} \ , \ \ \frac{\partial q_t}{\partial p}=\frac{\mathfrak{A}-\bar \mathfrak{A}}{2i} 
\end{equation}
and
\begin{equation}
\frac{\partial p_t}{\partial q}=\frac{\mathfrak{B}+\bar \mathfrak{B}}{2} \ , \ \ \frac{\partial p_t}{\partial p}=\frac{\mathfrak{B}-\bar \mathfrak{B}}{2i}
\end{equation}
and so we can express the canonical relations in terms of the variational forms.

For the canonical Poisson relations, $\{q_t,p_t\}=I$, we have 
\begin{equation}
\bar AB^{{\rm T}}-A\bar B^{{\rm T}}=2iI
\end{equation}
while for $\{q_t,q_t\}=0$ and $\{p_t,p_t\}=0$ we have 
\begin{equation}
\bar AA^{{\rm T}}-A\bar A^{{\rm T}}=0 \ , \ \ \bar BB^{{\rm T}}-B\bar B^{{\rm T}}=0
\end{equation}
by conjugation with the gauge unitary matrix. 

For the canonical Lagrange relations, $[\![q,p]\!]=I$, we have 
\begin{equation}
A^*B-B^*A=2iI
\end{equation}
by conjugation with the gauge unitary matrix, while $[\![q,q]\!]=0$ and $[\![p,p]\!]=0$ produce the adjoint of the above. 

Finally, beginning from $\bar A^{\rm T}B-\bar B^{\rm T}A=2iI$ and $Z=BA^{-1}$, we arrive at the relation 
\begin{eqnarray}
{\rm Im}\,Z=(A A^*)^{-1} \ .
\end{eqnarray}

Summarizing, the variational forms satisfy the following independent algebraic and dynamical relations, which hold along the flow 
\begin{eqnarray}
A^{\rm T}B-B^{\rm T}A=0 \ , \ \ AA^*\succ 0 \\ \nonumber
\bar AB^{\rm T}-A\bar B^{\rm T}=2iI \ , \ \ A\bar A^{\rm T}-\bar AA^{\rm T}=0 \ , \ \ B\bar B^{\rm T}-\bar BB^{\rm T}=0 \\ \nonumber 
A^*B-B^*A=2iI \\ \nonumber
{\rm Im}\,Z=(A A^*)^{-1} \ .
\end{eqnarray}

It is straightforward to incur that the antisymmetry of the Poisson and Lagrange brackets correspond to symmetry of the relations with respect to the interchange $A\leftrightarrow B$ along with complex conjugation, which bare the --dependent to the former-- identities 
\begin{eqnarray}
B^{\rm T}A-A^{\rm T}B=0 \ , \ \ BB^*\succ 0 \\ \nonumber
BA^*-\bar BA^{\rm T}=-2iI \ , \ \ \bar B B^{\rm T}- B\bar B^{\rm T}=0 \ , \ \ \bar AA^{\rm T}-A\bar A^{\rm T}=0 \\ \nonumber 
B^{{\rm T}}\bar A-A^{{\rm T}}\bar B=-2iI \\ \nonumber
{\rm Im}\,Z^{-1}=(BB^*)^{-1} \ .
\end{eqnarray}

As ${\rm Im}\,Z\succ 0$, the imaginary part of the anisotropy form $Ζ$ possesses a unique square root, by which that variational forms $A,B$ can be uniquely expressed
\begin{equation}
A=({\rm Im}\,Z)^{-1/2} \ , \ \ B=({\rm Im}\,Z^{-1})^{-1/2} \ .
\end{equation}


Utilizing the relations between the variational forms we simplify the expression of the semi-classical anisotropic Gaussian wave packet, $G^{Z,\hbar}_{(q,p)}$. Considering that 
\begin{eqnarray}
\fl \int\displaylimits_0^t{\rm tr}\Big(H_{pp}Z+H_{pq}\Big)\,d\tau=\int\displaylimits_0^t{\rm tr}\Big(H_{pp}BA^{-1}+H_{pq}AA^{-1}\Big)\,d\tau\\ \nonumber
=\int\displaylimits_0^t{\rm tr}\Big(\frac{d A}{d\tau}A^{-1}\Big)\,d\tau={\rm tr}\,\log\,A
\end{eqnarray}
we obtain the equivalent expression of the amplitude 
\begin{equation}
a(q,p,t)=\frac{1}{\sqrt{\det\,A(t)}}
\end{equation}
so that 
\begin{equation}
\fl G^{Z,\hbar}_{(q,p)}(x,t)= \frac{(\pi\hbar)^{-d/4}}{\sqrt{\det\,A(q,p,t)}}e^{\frac{i}{\hbar}\Big(\frac{p\cdot q}{2}+\mathcal{A}(q,p,t)+p_t\cdot (x-q_t)+\frac{1}{2}(x-q_t)\cdot Z(t)(x-q_t)\Big)}
\end{equation}
or 
\begin{equation}
\fl G^{Z,\hbar}_{(q,p)}(x,t)= (\pi\hbar)^{-d/4}\Big(\det\,{\rm Im}\,Z(t)\Big)^{1/4}e^{i\vartheta(t)}e^{\frac{i}{\hbar}\Big(\frac{p\cdot q}{2}+\mathcal{A}(q,p,t)+p_t\cdot (x-q_t)+\frac{1}{2}(x-q_t)\cdot Z(t)(x-q_t)\Big)}
\end{equation}
for a particular additional dynamical phase $\vartheta(t)$.

\subsection{The Anisotropic Gaussian Approximation for the Propagator}
\label{AGAP}

As $ G^Z _{(q,p)}$ is a semi-classical asymptotic solution of the Cauchy problem ($\ref{eq:ivp})$ with initial data $ \psi_0=G_{(q,p)}^\hbar$, we consider the following semi-classical approximation, in a certain sense, we \textit{define} the semi-classical flow by 
\begin{equation}
U^t_{sc} G_{(q,p)}^\hbar(x):=G^{Z,\hbar} _{(q,p)}(x,t) 
\end{equation}
by which we can construct a semi-classical asymptotic approximation of the phase space propagator
\begin{equation}
\mathcal{U}^t_{sc}=\mathcal{W} U_{sc}^t \mathcal{W}^{-1} \ .
\end{equation}

Concerning the validity of the approximation, the minimal time-scale which marks its limits is the semi-classical time-scale up to which $G^{Z,\hbar}_{(q,p)}$ remains a semi-classical solution of the given order and beyond which its Gaussian wave packet form, its micro-localization on the reference orbit on the Heisenberg scale in phase space, is irreversibly lost. This is the \textit{Ehrenfest time-scale} \cite{CTH1,SVT}, $T_E(\hbar)$, defined for the given base point, $(q,p)$, as
\begin{equation}
|\tilde g_H^{T_E(\hbar)}(q,p)|\asymp \frac{1}{\hbar^{1/2}} \ , \ \ \hbar\rightarrow 0^+
\end{equation}
where $\tilde g_H^t$ is the linearized Hamiltonian flow \cite{SVT}. The Ehrenfest time-scale is sensitive to the dynamical properties of the flow, be it global or local \cite{Fau1,Fau2,SVT}; in the case the flow is chaotic, in particular globally hyperbolic, it reads
\begin{equation}
T_E(\hbar)\asymp \log\,\hbar \ , \ \ \hbar\rightarrow 0^+ 
\end{equation}
while in the case the flow is completely integrable, it reads
\begin{equation}
T_E(\hbar)\asymp \frac{1}{\hbar^{1/2}} \ , \ \ \hbar\rightarrow 0^+ \ . 
\end{equation}

We, thus, conclude that, for a generic Hamiltonian function, one may safely assume the approximation for the propagation of a single Gaussian wave packet \cite{Rob1}
\begin{equation}
U^tG_{(q,p)}^\hbar(t)\sim G^{Z,\hbar}_{(q,p)}(t) \ , \ \ \hbar\rightarrow 0^+
\end{equation}
for 
\begin{equation}
t=o(\log\,\hbar) \ , \ \ \hbar\rightarrow 0^+ \ .
\end{equation}

Based on this approximation, the expression of the \textit{semi-classical asymptotic solution} of the Cauchy problem, $\psi^\hbar(t)$, is 
\begin{equation}
\fl \psi^\hbar(x,t)=\int K_{sc}(x,y,t;\hbar)\psi_0(y)\,dy=\Big(\frac{1}{2\pi\hbar}\Big)^{d/2}\int G^{Z,\hbar} _{(q,p)}(x,t)\Psi_0(q,p)\,dqdp \ .
\end{equation}

The corresponding kernel of the semi-classical propagator, $K_{sc}$, is expressed as 
\begin{equation}
K_{sc}(x,y,t;\hbar):=\Big(\frac{1}{2\pi\hbar}\Big)^{d}\int\bar G_{(q,p)}^\hbar(y) G^{Z,\hbar} _{(q,p)}(x,t)\,dqdp \ .
\end{equation}

We are interested in a semi-classical approximation for the phase space propagator. By substituting the approximate kernel $K_{sc}$ into the representation formulae ($\ref{eq:pswf}$) and ($\ref{eq:psprop}$), we obtain 
\begin{equation}
\mathcal{U}^t\Psi_0(q,p)\sim\mathcal{U}^t_{sc}\Psi_0(q,p) \ , \ \ \hbar\rightarrow 0^+
\end{equation}
defining the asymptotic solution for the phase space semi-classical problem 
\begin{eqnarray}
\fl \Psi^\hbar(q,p,t):=\int \mathcal{K}_{sc}(q,p,\eta,\xi,t;\hbar)\Psi_0(\eta,\xi)\,d\eta d\xi \\ \nonumber
=\Big(\frac{1}{2\pi\hbar}\Big)^{d}\int\!\!\int\bar G_{(q,p)}^\hbar(x) G^{Z,\hbar} _{(\eta,\xi)}(x,t)\Psi_0(\eta,\xi)\,dxd\eta d\xi \ .
\label{eq:sclsol}
\end{eqnarray}
We term this the \textit{Anisotropic Gaussian Approximation} for the propagator of the phase space Schr\"{o}dinger equation.

The kernel of the semi-classical Schr\"{o}dinger flow $\mathcal{U}^t_{sc}$ is 
\begin{eqnarray}
\mathcal{K}_{sc}(q,p,\eta,\xi,t;\hbar):=\Big(\frac{1}{2\pi\hbar}\Big)^{d}\int\bar G_{(q,p)}^\hbar(x) G^{Z,\hbar} _{(\eta,\xi)}(x,t)\,dx \ .
\end{eqnarray}

By introducing the Wick pseudo-complex co-ordinates $\eta-i\xi$ and its image under the Hamiltonian flow $\eta_t-i\xi_t$, by direct integration, we have 
\begin{eqnarray}
\fl \mathcal{K}_{sc}(q,p,\eta,\xi,t;\hbar)=\Big(\frac{1}{2\pi\hbar}\Big)^{d}\frac{2^{d/2}}{\sqrt{\det(A(\eta,\xi,t)-iB(\eta,\xi,t))}} \\ \nonumber 
\fl \times \exp\frac{i}{\hbar}\Bigg(\mathcal{A}(\eta,\xi,t)+\frac{\xi\cdot \eta-\xi_t\cdot \eta_t}{2}+\frac{1}{2}(q,p)\cdot J(\eta_t,\xi_t)+\frac{1}{2}\left(\begin{array}{ccc}
q-\eta_t\\
p-\xi_t 
\end{array}\right)^{{\rm T}}
Q
 \left(
\begin{array}{ccc}
q-\eta_t\\
p-\xi_t 
\end{array}\right)\Bigg)
\end{eqnarray}
or, equivalently,
\begin{eqnarray}
\fl \mathcal{K}_{sc}(q,p,\eta,\xi,t;\hbar)=\Big(\frac{1}{2\pi\hbar}\Big)^{d}\Big(\det\,\frac{\partial(\eta_t-i\xi_t)}{\partial(\eta-i\xi)}\Big)^{-1/2}\\ \nonumber
\fl \times \exp\frac{i}{\hbar}\Bigg(\mathcal{A}(\eta,\xi,t)+\frac{\xi\cdot \eta-\xi_t\cdot \eta_t}{2}+\frac{1}{2}(q,p)\cdot J(\eta_t,\xi_t)+\frac{1}{2}\left(\begin{array}{ccc}
q-\eta_t\\
p-\xi_t 
\end{array}\right)^{{\rm T}}
Q
 \left(
\begin{array}{ccc}
q-\eta_t\\
p-\xi_t 
\end{array}\right)\Bigg)
\label{eq:Lwp}
\end{eqnarray}
where the quadratic form
\begin{equation}
Q(\eta,\xi,t)=\left(
\begin{array}{ccc}
iI-i(I-iZ)^{-1} & \frac{1}{2}I-(I-iZ)^{-1}\\
\frac{1}{2}I-(I-iZ)^{-1} & i(I-iZ)^{-1} 
\end{array}\right)
\end{equation}
the \textit{double phase space anisotropy form}, is an element of Siegel upper half-space, $Q\in \Sigma_{2d}$ \cite{Fol}.

The semi-classical propagator does not define a unitary flow, but a semi-classically asymptotic unitary flow, so that as $\hbar\rightarrow 0^+$, in the appropriate weak sense, for $t,s\in\mathbb{R}$, 
\begin{equation}
U_{sc}^t U_{sc}^{s}\sim U_{sc}^{t+s} \ , \ \ \hbar\rightarrow 0^+
\end{equation}
and
\begin{equation}
(U_{sc}^t)^*\sim (U_{sc}^{t})^{-1}\sim U_{sc}^{-t} \ , \ \ \hbar\rightarrow 0^+ \ .
\end{equation}

The semi-classical phase space propagator, $\mathcal{U}^t_{sc}$, defines semi-classically unitary Schr\"{o}dinger flow in $\mathfrak{F}$, as it preserves the Fock-Bargmann analyticity constraints (\ref{eq:FBc}), while for $t,s\in\mathbb{R}$, 
\begin{equation}
\mathcal{U}_{sc}^t \mathcal{U}^{s}_{sc}\sim \mathcal{U}_{sc}^{t+s} \ , \ \ \hbar\rightarrow 0^+ \ .
\end{equation}
and
\begin{equation}
(\mathcal{U}_{sc}^t)^*\sim (\mathcal{U}_{sc}^{t})^{-1}\sim \mathcal{U}_{sc}^{-t} \ , \ \ \hbar\rightarrow 0^+ \ .
\end{equation}

At $t=0$, the semi-classical wave packet phase space propagator shares the reproducing property of the exact phase space propagator,
\begin{eqnarray} 
\fl \mathcal{K}_{sc} (q,p,\eta,\xi,0;\hbar)=\mathcal{K} (q,p,\eta,\xi,0;\hbar)=\Big(\frac{1}{2\pi\hbar}\Big)^{d}\int\bar G_{(q,p)}^\hbar(x) G^{Z,\hbar} _{(\eta,\xi)}(x,0)\,dx \\ \nonumber
=\Big(\frac{1}{2\pi\hbar}\Big)^{d}\int\bar G_{(q,p)}^\hbar(x) G_{(\eta,\xi)}^\hbar(x)\,dx=b(q,p,\eta,\xi;\hbar)
\end{eqnarray}
the Bergmann repoducing kernel.

\subsection{Relation to Initial Value Representations and the Littlejohn Approximation}

In \cite{Lit1}, Littlejohn constructed a semi-classical phase space propagator based on the Nearby Orbit Approximation, as an explicit action of Weyl shifts and metaplectic operators, generalizing the approximation for the dynamics of Liouville densities in the quantum mechanical framework. 

This construction, however, does not fall into the scheme of Initial Value Representations. The Littlejohn propagator utilizes a single reference orbit, emanating from a given phase space point on which the initial data $\psi_0 $ is assumed to be centered at and localized, without necessarily being Gaussian. 

The semi-classical Littlejohn flow reads \cite{Lit2}
\begin{equation}
\mathcal{U}_L^t\Psi(q,p):=\int \mathcal{K}_L(q,p,\eta,\xi,t;\hbar)\Psi(\eta,\xi)\,d\eta d\xi
\end{equation}
where the kernel is 
\begin{equation}
\fl \mathcal{K}_L(q,p,\eta,\xi,t;\hbar)=\Big(\frac{1}{2\pi\hbar}\Big)^d\int \bar G^\hbar _{(q,p)}(x)e^{\frac{i}{\hbar}\vartheta(\eta,\xi,t)}(\mathcal{T}_{g^t(\eta,\xi)}^\hbar M(\eta,\xi,t)G_0^\hbar)(x)\,dx
\end{equation}
where $\mathcal{T}_{(q,p)}^\hbar$ is the Weyl shift \cite{CoRo} and $M$ is a metaplectic operator whose dynamics is governed by
\begin{equation}
\frac{dM}{dt}=\frac{i\hbar}{2}\Big(\eta\cdot H_{\eta\eta}\eta+\frac{\partial}{\partial\eta}\cdot H_{\xi\eta}\eta+\eta\cdot H_{\eta\xi}\frac{\partial}{\partial\eta}+\frac{\partial}{\partial\eta}\cdot H_{\xi\xi}\frac{\partial}{\partial\eta}\Big)M
\end{equation}
with initial condition 
\begin{equation}
M(0)={\rm Id}
\end{equation}
the Hessian elements evaluated along the flow, at $(\eta_t,\xi_t)$, while the additional phase is $\vartheta(\eta,\xi,t)=\mathcal{A}(\eta,\xi ,t)-\frac{\xi_t\cdot \eta_t}{2}$.

The propagator fixes the initial state from $(q,p)$ to the origin, symplectically `rotates' it in phase space by the action of the metaplectic operator $M(t)$, and shifts it along the reference orbit modulating by adding the action phase.

A different approach, on the lines of which we constructed the semi-classical phase space propagator, is that of Initial Value Representations for solutions of the Schr\"{o}dinger equation. By beginning with the completeness relation of the isotropic Gaussian states in phase space $(\ref{eq:completeness})$,
\begin{equation}
K(x,y,t;\hbar)=\Big(\frac{1}{2\pi\hbar}\Big)^{d} \int U^tG_{(q,p)}^\hbar(x)\bar G_{(q,p)}^\hbar(y)\,dqdp 
\end{equation}
where $K$ is the kernel of the propagator, the following approximate phase space resolutions are usually made in this context, as $\hbar\rightarrow 0^+$,
\begin{eqnarray}
\fl K(x,y,t;\hbar)\sim \Big(\frac{1}{2\pi\hbar}\Big)^{d} \int G^{Z,\hbar}_{(q,p)}(x,t)\bar G_{(q,p)}^\hbar(y)\,dqdp  \\ \nonumber
\fl K(x,y,t;\hbar)\sim \Big(\frac{1}{2\pi\hbar}\Big)^{d} \int e^{\frac{i}{\hbar}S(q,p,t)}G_{(q,p)}^\hbar(x)\bar G_{(q,p)}^\hbar(y)\,dqdp \\ \nonumber 
\fl K(x,y,t;\hbar)\sim \Big(\frac{1}{2\pi\hbar}\Big)^{d} \int c(q,p,t;\hbar)e^{\frac{i}{\hbar}S(q,p,t)}G_{(q,p)}^\hbar(x)\bar G_{(q,p)}^\hbar(y)\, dqdp  
\end{eqnarray}
the above being, respectively, the \textit{Thawed} or \textit{Anisotropic Gaussian Approximation}, the \textit{Frozen Gaussian Approximation}, and the \textit{Herman-Kluk Approximation} \cite{HeKl,Rob2,RoSw}.

\section{Semi-Classical Time Evolution in Phase Space}
\label{solutions}

We now turn to the issue of central interest, the semi-classical Cauchy problem for the Weyl-symmetrized phase space Schr\"{o}dinger equation, 
\begin{equation}
\Big(i\hbar\, \frac{\partial }{\partial t}-\widecheck H\Big)\Psi(t)=0 \ , \ \ t\in[0,T]
\end{equation}
with initial condition 
\begin{equation}
\Psi(0)=\Psi_0
\end{equation}
for some semi-classical inital state $\Psi_0\in\mathfrak{F}$.

We construct an asymptotic propagator and asymptotic solution of the problem on the basis of the Anisotropic Gaussian Approximation, by semi-classically evolving semi-classical phase space states under the semi-classical Schr\"{o}dinger flow, $\mathcal{U}^t_{sc}$.

As was noted in sub-section \ref{AGAP}, the approximation is based on the corresponding approximation for single Gaussian wave packet dynamics, which is valid for time-scales well shorter than the Ehrenfest time-scale, $T=o(\log\,\hbar)$. The issue of whether the the Anisotropic Gaussian Approximation is valid for time-scales well beyond the Ehrenfest time, for the wave packet superposition in configuration space or in phase space, is an open one.

\subsection{Semi-Classical Asymptotics of the Phase Space Propagator}

A way of probing generic properties of semi-classical time evolution in phase space generated by a given Hamiltonian operator, independent of specific initial state, is by considering the relevant asymptotics of the phase space propagator, $\mathcal{U}^t$.

We begin by considering the time evolution of the propagator in configuration space, $U^t$, as in (\ref{eq:eqprop}),  
\begin{equation}
\Big(i\hbar\,\frac{d}{dt}-\widehat H \Big)U^t=0 \ , \ \ t\in [0,T]
\end{equation}
with initial condition
\begin{equation}
U^0={\rm Id} \ .
\end{equation}
Following Berezin and Shubin \cite{BeSh}, we consider its Weyl symbol, $\sigma_W(U^t)$, and the dynamics the above equation of motion induces on it. Using the composition formula for Weyl operators, we may express the evolution for the Weyl symbol $\sigma_W(U^t)(q,p;\hbar)$, given that $\sigma_W(\widecheck H)(q,p)=H$, as 
\begin{equation}
i\hbar\,\frac{\partial\sigma_W(U^t)}{\partial t}=H\ast_\hbar\sigma_W(U^t) \ , \ \ t\in [0,T]
\end{equation}
or, giving the explicit form of the deformed non-commutative product $\ast_\hbar$ \cite{Ber,BeSh,Wei}
\begin{equation}
i\hbar\frac{\partial\sigma_W(U^t)}{\partial t}=\sum_{\alpha,\beta\in\mathbb{N}^d_0}\Big(\frac{i\hbar}{2}\Big)^{|\alpha+\beta|}\frac{(-1)^{|\alpha|}}{\alpha!\beta!}\frac{\partial^{\alpha+\beta}H}{\partial q^\alpha\partial p^\beta}(q,p)\,\frac{\partial^{\alpha+\beta}\sigma_W(U^t)}{\partial p^\alpha\partial q^\beta}
\end{equation}
with initial condition
\begin{equation}
\sigma_W(U^0)(q,p;\hbar)=1 \ .
\end{equation}

Following Berezin and Shubin\footnote{We note that Berezin and Shubin, as well as Marinov, consider the asymptotics of the adjoint of the propagator, hence the opposite sign for the phase function $\phi$ in the symmetrized Hamilton-Jacobi and transport equations.}, we consider the WKB approximation for the symbol 
\begin{equation}
\sigma_W(U^t)(q,p;\hbar)\sim \chi(q,p,t;\hbar)\,e^{\frac{i}{\hbar}\phi(q,p,t)} \ , \ \ \hbar\rightarrow 0^+
\end{equation}
where
\begin{equation}
\chi(q,p,t;\hbar)\sim\sum_{r=0}^\infty \hbar^r \chi_r(q,p,t) \ , \ \ \hbar\rightarrow 0^+ \ .
\end{equation}

The above WKB approximation leads to the well known Berezin-Shubin-Marinov Hamilton-Jacobi equation \cite{Alm,BeSh,Mar}
\begin{equation}
\frac{\partial\phi}{\partial t}+H\Big(q+\frac{1}{2}\frac{\partial\phi}{\partial p},p-\frac{1}{2}\frac{\partial\phi}{\partial q}\Big)=0
\end{equation}
along with a corresponding transport equation for the leading term amplitude (here we use the notation $\chi\equiv \chi_0$)
\begin{equation}
\fl \frac{\partial \chi}{\partial t}-\Big(\frac{\partial H}{\partial \eta}\cdot \frac{\partial \chi}{\partial \xi}-\frac{\partial H}{\partial \xi}\cdot \frac{\partial \chi}{\partial \eta}\Big)+\frac{1}{2}{\rm tr}\,\Big(H_{\eta\eta}\,\frac{\partial^2 \phi}{\partial\xi^2}+2\,H_{\eta\xi}\,\frac{\partial^2 \phi}{\partial p\partial q}+H_{\xi\xi}\,\frac{\partial^2 \phi}{\partial q^2}\Big)\,\chi=0
\end{equation}
where 
\begin{equation}
\eta=q+\frac{1}{2}\frac{\partial\phi}{\partial p} \ , \ \ \xi=p-\frac{1}{2}\frac{\partial\phi}{\partial q}
\end{equation}
along with initial conditions 
\begin{equation}
\phi(q,p,0)=0 \ , \ \ \chi(q,p,0)=1 \ .
\end{equation}

Berezin and Shubin \cite{BeSh} introduced the above canonical system  in considering the asymptotics of the symbol of the Schr\"{o}dinger propagator in the Weyl quantization, standing as a semi-classical correction to the Hamiltonian flow as canonical transformation and the Liouville transport equation; subsequently, Marinov \cite{Mar} developed a purely classical geometric argument for the deduction of the same variant of the Hamilton-Jacobi equation as a generator of canonical transformations.

Using the relations between the kernel $\mathcal{K}$ and Weyl symbol of the propagator $U^t$
\begin{eqnarray}
\fl \mathcal{K}(q,p,\eta,\xi,t;\hbar)=\Big(\frac{1}{2\pi\hbar}\Big)^{2d}\int\!\!\int\!\!\int \bar G^\hbar_{(q,p)}(x)G_{(\eta,\xi)}^\hbar(y)\\ \nonumber 
\times e^{\frac{i}{\hbar}(x-y)\cdot v}\sigma_W(U^t)\Big(\frac{x+y}{2},\xi;\hbar\Big)\,dxdydv
\end{eqnarray}
we deduce the induced WKB semi-classical approximation for the kernel $\mathcal{K}$
\begin{equation}
\mathcal{K}(q,p,\eta,\xi,t;\hbar)\sim \Big(\frac{1}{2\pi\hbar}\Big)^d\varphi(q,p,\eta,\xi,t;\hbar)\,e^{\frac{i}{\hbar}F(q,p,\eta,\xi,t)} \ , \ \ \hbar\rightarrow 0^+
\end{equation}
where 
\begin{equation}
\varphi(q,p,\eta,\xi,t;\hbar)\sim\sum_{r=0}^\infty \hbar^{r} \varphi_r(q,p,\eta,\xi,t) \ , \ \ \hbar\rightarrow 0^+ 
\label{eq:asK}
\end{equation}
noting that $F$ is necessarily complex valued with non-negative imaginary part.

The semi-classical dynamics of the kernel induced by the dynamics of the symbol yield variant of the symmetrized Hamilton-Jacobi equation for  the phase
\begin{equation}
\frac{\partial F}{\partial t}+H\Big(\frac{q}{2}-\frac{\partial F}{\partial p},\frac{p}{2}+\frac{\partial F}{\partial q}\Big)=0
\end{equation}
and a variant transport equation (here we use the notation $\varphi\equiv \varphi_0$)
\begin{equation}
\frac{\partial \varphi}{\partial t}-\frac{1}{2}\,{\rm tr}\Bigg(H_{qq}\frac{\partial^2 F}{\partial q^2}+H_{pq}\frac{\partial^2 F}{\partial p\partial q}+H_{qp}\frac{\partial^2 F}{\partial q\partial p}+H_{pp}\frac{\partial^2 F}{\partial p^2}\Bigg)\,\varphi=0
\end{equation}
along with initial conditions 
\begin{equation}
\fl F(q,p,\eta,\xi,0)=\frac{1}{2}(q,p)\cdot J(\eta,\xi) +\frac{i}{4}\Big(|q-\eta|^2+|p-\xi|^2\Big)\ , \ \ \varphi(q,p,\eta,\xi,0)=1
\end{equation}
so that 
\begin{equation}
\mathcal{K}(q,p,\eta,\xi,0;\hbar)=b(q,p,\eta,\xi;\hbar)=\Big(\frac{1}{2\pi\hbar}\Big)^d\langle G^\hbar_{(q,p)},G^\hbar_{(\eta,\xi)}\rangle 
\end{equation}
the Bergmann repoducing kernel (see \ref{WPTr}). 

We arrive at a similarly structured canonical system, which seems to inherit --under certain dilations-- the Weyl symmetrization, the mportance of which, on the level of symplectic geometry, was pointed out by de Gosson \cite{Gos1,Gos2}.

The technical difficulty in the phase space asymptotics is the inescapably from complex valued dynamics, for example, for the symmetrized Hamilton-Jacobi equation and transport equations in the case the Hamiltonian is not analytic. This demands the use of methods of the complex WKB theory.

\subsection{Posing of the Semi-Classical Cauchy Problem}

We consider $\Psi_0$ to be the phase space image of the WKB state under $\mathcal{W}$, $\Psi_0=\mathcal{W}\psi_0$,
\begin{equation}
\label{eq:WKB}
\psi_0(x)=R_0(x)e^{\frac{i}{\hbar}S_0(x)}
\end{equation}
where $S_0\in C^\infty (\mathbb{R}^d,\mathbb{R})$ and $R_0\in \mathcal{S}(\mathbb{R}^d,\mathbb{R})$, so that $\int R^2_0\, dx=1$ and ${\rm det}\,S_0''(x)\neq 0$ everywhere.

By the complex stationary phase theorem (see \ref{CSPT}), we determine the initial WKB data in phase space, $\Psi_0=\mathcal{W}\psi_0$, as \cite{NSS} 
\begin{eqnarray}
\fl \Psi_0(q,p)=(\pi\hbar)^{-d/4}\frac{{}^r \! R_0(z(q,p))}{\sqrt{\det\Big(I-i \,{}^r \! S''_0(z(q,p))\Big)}}\\ \nonumber 
\fl \times \exp\frac{i}{\hbar}\Bigg(\,{}^r \! S_0(z(q,p))-p\cdot \Big(z(q,p)-q\Big)+\frac{i}{2}\Big(z(q,p)-q\Big)^2-\frac{p\cdot q}{2}\Bigg)\Big(1+o(\hbar)\Big) 
\label{eq:wkb0}
\end{eqnarray}
where ${}^r \! S_0$ and $\,{}^r \! R_0$ stand for the $r$-analytic extension (see \ref{CSPT}) to the complex variable $z=x+iy\in\mathbb{C}^d$ of $S_0,R_0$ respectively, for any $s\in\mathbb{N}$, and $z=z(q,p)$ stands for the sole complex solution of the stationary equation 
\begin{equation}
\frac{\partial \,{}^r \! S_0}{\partial z}(z)-p+i(z-q)=0
\end{equation}
(see \ref{CSPT}). The solution is given explicitly 
\begin{equation}
z(q,p)=q+i\Big(I-iS_0''(q)\Big)^{-1}\Big(S_0'(q)-p\Big)
\end{equation}
which, on the Lagrangian manifold induced by the initial data
\begin{equation}
\Lambda_0:=\{(q,p)\in\mathbb{R}^{2d}|\,p=\frac{\partial S_0}{\partial q}(q)\}
\end{equation}
reduces to the simple real expression
\begin{equation}
z|_{\Lambda_0}(q,p)=q \ .
\end{equation}
We stress that the solution is real only on the Lagrangian manifold $\Lambda_0$.

The imaginary part of the phase of the initial data vanishes along the Lagrangian manifold, $(q,p)\in\Lambda_0$, so that 
\begin{equation}
\fl \Psi_0|_{\Lambda_0}(q,p)=(\pi\hbar)^{-d/4}\frac{R_0(q)}{\sqrt{\det\,\Big(I-iS''_0(q)\Big)}}\,e^{\frac{i}{\hbar}(-\frac{p\cdot q}{2}+S_0(q))}\Big(1+o(\hbar)\Big) \ .
\end{equation}

We proceed to construct tha asymptotic solution of the Cauchy problem for the phase space Schr\"{o}dinger equation, by applying the semi-classical phase space Schr\"{o}dinger flow on the phase space WKB initial data,
\begin{equation}
\Psi^\hbar(q,p,t)=\int\mathcal{K}_{sc}(q,p,\eta,\xi,t;\hbar)\Psi_0(\eta,\xi)\,d\eta d\xi 
\end{equation}
where we define 
\begin{eqnarray}
\fl F(q,p,\eta,\xi,t)=\,{}^r \! S_0(z(\eta,\xi))-\xi\cdot \Big(z(\eta,\xi)-\eta\Big)+\frac{i}{2}\Big(z(\eta,\xi)-\eta\Big)^2+\mathcal{A}(\eta,\xi,t)\\ \nonumber 
\fl -\frac{\xi_t\cdot \eta_t}{2}+\frac{1}{2}(q,p)\cdot J(\eta_t,\xi_t)+\frac{1}{2}\left(\begin{array}{ccc}
q-\eta_t\\
p-\xi_t 
\end{array}\right)^{{\rm T}}
Q(\eta,\xi,t)
 \left(
\begin{array}{ccc}
q-\eta_t\\
p-\xi_t 
\end{array}\right)
\end{eqnarray}
and 
\begin{eqnarray}
\fl \varphi(\eta,\xi,t;\hbar)=(\pi\hbar)^{-d/4}\Big(\det\,\frac{\partial(\eta_t-i\xi_t)}{\partial(\eta-i\xi)}\Big)^{-1/2}\frac{{}^r \! R_0(z(\eta,\xi))}{\sqrt{\det\Big(I-i \,{}^r \! S''_0(z(\eta,\xi))\Big)}}
\end{eqnarray}
by which we reach a semi-classical Fourier integral representation of the semi-classical asymptotic solution
\begin{equation}
\Psi^\hbar(q,p,t)=\Big(\frac{1}{2\pi\hbar}\Big)^d\int\varphi(\eta,\xi,t;\hbar)e^{\frac{i}{\hbar}F(q,p,\eta,\xi,t)}\,d\eta d\xi \ .
\end{equation}

\subsection{The Canonical System in Double Phase Space}

In the following we closely follow the methodology of Maslov \cite{Mas2}.

We begin with the semi-classical Fourier integral representation of the asymptotic solution 
\begin{equation}
\Psi^\hbar(q,p,t)=\Big(\frac{1}{2\pi\hbar}\Big)^d\int \varphi(\eta,\xi,t;\hbar)e^{\frac{i}{\hbar}F(q,p,\eta,\xi)}\,d\eta d\xi \ .
\end{equation}
By inserting the integral representation in the phase space Schr\"{o}dinger and demanding it to be an order $O(\hbar)$ asymptotic solution, we obtain
\begin{equation}
\fl \Big(i\hbar\,\frac{\partial }{\partial t}-\widecheck H \Big)\Psi^\hbar(q,p,t)=\Big(\frac{1}{2\pi\hbar}\Big)^d\int\Big(i\hbar\,\frac{\partial }{\partial t}-\widecheck H \Big)\varphi(\eta,\xi,t;\hbar)\,e^{\frac{i}{\hbar}F(q,p,\eta,\xi,t)}\,d\eta d\xi
\end{equation}
where the operator $\widecheck H $ acts with respect to $(q,p)$, and thus
\begin{equation}
\fl \Big(i\hbar\,\frac{\partial }{\partial t}-\widecheck H \Big)\Psi^\hbar=\Big(\frac{1}{2\pi\hbar}\Big)^d\int\Big(i\hbar\,\varphi^{-1}\frac{\partial \varphi}{\partial t}-\frac{\partial F }{\partial t}-e^{-\frac{i}{\hbar}F }\widecheck H e^{\frac{i}{\hbar}F}\Big)\varphi \,e^{\frac{i}{\hbar}F }\,d\eta d\xi \ .
\end{equation}

By the semi-classical commutation formula (see \ref{CSPT}), for $X=(q,p)$,
\begin{eqnarray}
\fl H\Big(\stackrel{\boldsymbol{\omega}}{\frac{X}{2}}+i\hbar\, J\stackrel{\boldsymbol{\omega}}{\frac{\partial}{\partial X}}\Big)e^{\frac{i}{\hbar}F(X)}\sim e^{\frac{i}{\hbar}F(X)}\Bigg(\,{}^s\! H\Big(\frac{X}{2}-J\frac{\partial F}{\partial X}(X)\Big)\\ \nonumber 
+\frac{i\hbar}{2}\, {\rm tr}\,\frac{\partial^2 F}{\partial X^2} \,{}^{s }\!\Big(\frac{\partial^2 H}{\partial X^2}\Big)\Big(\frac{X}{2}-J\frac{\partial F}{\partial X}(X)\Big)\Bigg) \ , \ \ \hbar\rightarrow 0^+ \ .
\end{eqnarray}
we obtain the Weyl-symmetrized Hamilton-Jacobi equation for the phase space eikonal equation 
\begin{equation}
\frac{\partial F }{\partial t}+{}^s\!H\Big(\frac{q}{2}-\frac{\partial F }{\partial p},\frac{p}{2}+\frac{\partial F }{\partial q}\Big)=0
\label{eq:HJeq}
\end{equation} 
and the phase space transport equation 
\begin{equation}
\frac{\partial \varphi}{\partial t}-\frac{1}{2}\,{\rm tr}\Bigg(\frac{\partial^2 F}{\partial X^2}\,{}^s\!\Big(\frac{\partial^2 H}{\partial X^2}\Big)\Big(\frac{X}{2}-J\frac{\partial F}{\partial X}\Big)\Bigg)\,\varphi=0
\label{eq:TReq}
\end{equation}
where the Hessian elements are evaluated at the phase space points $\Big(\frac{X}{2}-J\frac{\partial F }{\partial X}\Big)=\Big(\frac{q}{2}-\frac{\partial F }{\partial p},\frac{p}{2}+\frac{\partial F }{\partial q}\Big)$. The above are accompanied by the corresponding initial conditions for \textit{complex valued} initial data\footnote{$\varphi_0$ here is not to be confused with the notation used in (\ref{eq:asK}).}
\begin{equation}
F(q,p,\eta,\xi,0)=:F_0(q,p,\eta,\xi) \ , \ \ \varphi(\eta,\xi,0;\hbar)=:\varphi_0(\eta,\xi;\hbar)
\end{equation}
where 
\begin{eqnarray}
\fl F_0(q,p,\eta,\xi)={}^r\! S_0(z(\eta,\xi))-\xi\cdot \Big(z(\eta,\xi)-\eta\Big)+\frac{i}{2}\Big(z(\eta,\xi)-\eta\Big)^2-\frac{1}{2}\xi\cdot \eta\\ \nonumber
+\frac{1}{2}(q,p)\cdot J(\eta,\xi)+\frac{i}{4}\Big(|q-\eta|^2+|p-\xi|^2\Big)
\end{eqnarray}
and
\begin{equation}
\varphi_0(\eta,\xi;\hbar)=(\pi\hbar)^{-d/4}\,\frac{{}^r\!R_0(z(\eta,\xi))}{\sqrt{{\rm det}\,\Big(I-i\,{}^r\!S_0''(z(\eta,\xi))\Big)}} \ .
\end{equation}
We note that the canonical system is parametrized by the integration variable $Y=(\eta,\xi)$.

The imaginary part of the initial phase is non-negative, ${\rm Im}\,F_0(X,Y)\geq 0$ everywhere, as, by the complex stationary phase theorem (see \ref{CSPT}) there exists $c>0$ such that 
\begin{equation}
\fl {\rm Im}\,\Bigg({}^r\! S_0(z(\eta,\xi))-\xi\cdot \Big(z(\eta,\xi)-\eta\Big)+\frac{i}{2}\Big(z(\eta,\xi)-\eta\Big)^2\Bigg)\geq c\,|{\rm Im}\,z(\eta,\xi)|^2
\end{equation}
while it becomes zero only for $z(\eta,\xi)=\eta$, which happens for $(\eta,\xi)\in\Lambda_0$, and $(q,p)=(\eta,\xi)$. 

Focusing on the dynamic variable of the canonical system, $X=(q,p)$, we term the zero-level set of ${\rm Im}\,F_0$ by $\delta_{Y}^0$, 
\begin{equation}
\delta_{Y}^0=\{X\in\mathbb{R}^{2d}|\,{\rm Im}\,F_0(X,Y)=0\}
\end{equation}
so that $\delta_{Y}^0=\Lambda_0$ if $Y\in\Lambda_0$, otherwise $\delta_{Y,0}=\emptyset$, where 
\begin{equation}
\Lambda_0:=\{(q,p)\in\mathbb{R}^{2d}|\,p=\frac{\partial S_0}{\partial q}(q)\} \ .
\end{equation}

When $Y\in\Lambda_0$, the set $\delta_{Y}^0$ is a smooth isotropic sub-manifold of maximal dimension, ${\rm dim}\,\delta_Y^0=d$, of the phase space $\mathbb{R}^{2d}$, i.e., a smooth Lagrangian sub-manifold. For connected open sub-sets $U\subset \delta_Y^0$ we define systems of local co-ordinates $(U,\alpha)$, where $\alpha=(\alpha_1,\ldots, \alpha_d)$, such that the sub-manifold $\delta_Y^0$ has the local parametrization
\begin{equation}
X=X_0(\alpha)=(q_0(\alpha),p_0(\alpha))=\Big(q_0(\alpha),\frac{\partial S_0}{\partial q}(q_0(\alpha))\Big) \ .
\end{equation}

For $t>0$, the imaginary part
\begin{eqnarray}
\fl {\rm Im}\,F(q,p,\eta,\xi,t)={\rm Im}\,\Bigg(\,{}^r \! S_0(z(\eta,\xi))-\xi\cdot \Big(z(\eta,\xi)-\eta\Big)+\frac{i}{2}\Big(z(\eta,\xi)-\eta\Big)^2\Bigg)\\ \nonumber 
+\frac{1}{2}\left(\begin{array}{ccc}
q-\eta_t\\
p-\xi_t 
\end{array}\right)^{{\rm T}}
{\rm Im}\,Q(\eta,\xi,t)
 \left(
\begin{array}{ccc}
q-\eta_t\\
p-\xi_t 
\end{array}\right)
\end{eqnarray}
remains non-negative, ${\rm Im}\,F(q,p,\eta,\xi,t)\geq 0$ everywhere, taking into account that ${\rm Im}\,Q(\eta,\xi,t)\succ 0$. 

The zero level-set of the imaginary part of the phase is 
\begin{equation}
\delta_Y=\{(X,t)\in\mathbb{R}^{2d+1}|\,{\rm Im}\,F(X,Y,t)=0\}
\end{equation}
where $\delta_Y= \bigcup_{t\geq 0}\Lambda_{t}\times\{t\}$ if $Y\in\Lambda_0$, otherwise $\delta_Y=\emptyset$, where 
\begin{equation}
\Lambda_t:=g^t_H\Lambda_0=\{(q,p)\in\mathbb{R}^{2d}|\,p=\frac{\partial S}{\partial q}(q,t)\}
\end{equation}
the Lagrangian sub-manifold of the solution of the Cauchy problem for the Hamilton-Jacobi equation 
\begin{equation}
\frac{\partial S}{\partial t}+H\Big(x,\frac{\partial S}{\partial x}\Big)=0 \ , \ \ t\in [0,T]
\end{equation}
with initial condition 
\begin{equation}
S(x,0)=S_0(x) \ .
\end{equation}

When $Y\in\Lambda_0$, the set $\delta_{Y}$ is a smooth sub-manifold of dimension ${\rm dim}\,\delta_Y=d+1$, of the extended phase space $\mathbb{R}^{2d+1}$, while $\partial\delta_Y=\delta_Y^0$.

For connected open sub-sets $V\subset \delta_Y$ we define systems of local co-ordinates $(V,(\alpha,t))$, where $(\alpha,t)=(\alpha_1,\ldots, \alpha_d,t)$, such that the sub-manifold $\delta_Y$ has the local parametrization
\begin{equation}
(X,t)=(X_t(\alpha),t)=\Big((q_t(\alpha),p_t(\alpha)),t\Big)=\Big(q_t(\alpha),\frac{\partial S}{\partial q}(q_t(\alpha),t),t\Big) \ .
\end{equation}

\subsection{The Characteristic System in Double Phase Space}

The characteristic system of the canonical system induces a Hamiltonian flow on the double phase space $\mathbb{R}^{4d}$ with Hamiltonian function the symbol $\sigma_W(\widecheck H)$ and initial data in $\delta_Y^0$, where $Y\in\Lambda_0$. We define the Lagrangian sub-manifold in double phase space
\begin{equation}
\Lambda_{Y,0}=\{(X,P)\in\mathbb{R}^{4d}|\,P=\frac{\partial F_0}{\partial X}(X,Y)\}
\end{equation}
as well as the propagated Lagrangian sub-manifold $\Lambda_{Y,t}=g^t_H\Lambda_{Y,0}$.

In particular, we have 
\begin{equation}
\frac{dX}{dt}=\frac{\partial\sigma_W(\widecheck H)}{\partial P}(X,P) \ , \ \ \frac{dP}{dt}=-\frac{\partial\sigma_W(\widecheck H)}{\partial X}(X,P)
\end{equation}
with initial conditions
\begin{equation}
X|_{t=0}=X_0(\alpha) \ , \ \ P|_{t=0}=P_0(\alpha)=\frac{\partial F_0}{\partial X}(X_0(\alpha),X_0(\alpha))
\end{equation}
where $(U,\alpha)$ is a system of local co-ordinates in $\Lambda_0$, with $\alpha=(\alpha_1,\ldots, \alpha_d)$. The system has a unique solution in a given neighborhood $U\subset \Lambda_0$ and $\alpha\in U$, for times beyond $T$,
\begin{equation}
(X,P)=(X_t(\alpha),P_t(\alpha)) \ , \ \ t\geq 0 \ .
\end{equation}
We note that 
\begin{equation}
P_0(\alpha)=\frac{\partial F_0}{\partial X}(X_0(\alpha),X_0(\alpha))=\frac{1}{2}JX_0(\alpha)
\end{equation} 
so that the initial data are real, and thus the dynamics is real and unique.

The form of the equations of Hamilton can be simplified
\begin{equation}
\frac{dX}{dt}=J\frac{\partial H}{\partial X}\Big(\frac{X}{2}-JP\Big) \ , \ \ \frac{dP}{dt}=-\frac{1}{2}\frac{\partial H}{\partial X}\Big(\frac{X}{2}-JP\Big)
\end{equation}
from where we deduce the $2d$ integrals of the motion
\begin{equation}
c_k(X,P)=\frac{X_k}{2}+\sum_{l=1}^{2d}J_{kl}P_l \ , \ \ k=1,\ldots, 2d
\end{equation}
which take the value zero due to the condition $P_0(\alpha)=\frac{1}{2}JX_0(\alpha)$,
\begin{equation}
c(\alpha)=\frac{X_0(\alpha) }{2}+JP_0(\alpha)=0 \ .
\end{equation}
Since $c=0$, we have $\frac{X}{2}-JP=X$, and so the equations of Hamilton are further simplified to
\begin{equation}
\frac{dX}{dt}=J\frac{\partial H}{\partial X}(X) \ , \ \ \frac{dP}{dt}=-\frac{1}{2}\frac{\partial H}{\partial X}(X) \ .
\end{equation}

The canonical projection of the Hamiltonian flow with respect to $\sigma_W(\widecheck H)$ in $\mathbb{R}^{4d}$ onto $\mathbb{R}^{2d}$ is 
in fact a Hamiltonian flow with Hamiltonian $H$ in $\mathbb{R}^{2d}$
\begin{equation}
\frac{dX}{dt}=J\frac{\partial H}{\partial X}(X)
\end{equation}
with initial condition 
\begin{equation}
X|_{t=0}=X_0(\alpha) \ .
\end{equation}
The projected flow is explicitly given by 
\begin{equation}
X_t(\alpha)=g^t_HX_0(\alpha)
\end{equation}
meaning $(q_t(\alpha),p_t(\alpha))=g^t_H(q_0(\alpha),p_0(\alpha))$. Finally, the solution of the characteristic systems is given by 
\begin{equation}
(X,P)=g_{\sigma_W(\widecheck H)}^t(X_0(\alpha),P_0(\alpha))=\Big(X_t(\alpha),\frac{1}{2}JX_t(\alpha)\Big) \ .
\end{equation}

We denote the image of the orbit of duration $t$ emanating from point $(X_0(\alpha),P_0(\alpha))$ by 
\begin{equation}
\Gamma^t(X_0(\alpha),P_0(\alpha))=\bigcup_{0\leq\tau\leq t}\{g^\tau_{\sigma_W(\widecheck H)}(X_0(\alpha),P_0(\alpha))\}\subset \mathbb{R}^{4d} \ .
\end{equation}
 
If $\Pi:\mathbb{R}^{4d}\rightarrow \mathbb{R}^{2d}$ is the canonical projection form the double phase space onto the phase space, then
\begin{equation}
\gamma^tX_0(\alpha)=\Pi\Gamma^t(X_0(\alpha),P_0(\alpha)) \ .
\end{equation}

Central to the structure of the characteristic system is the symplectic sub-space 
\begin{equation}
L=\{(X,P)\in\mathbb{R}^{4d}|\,P=\frac{1}{2}JX\}
\end{equation}
with symplectic form $\Omega^2|_L=dP\wedge dX|_L=\omega^2=dp\wedge dq$, which is an invariant sub-manifold of the flow 
\begin{equation}
g^t_{\sigma_W(\widecheck H)}L=L \ .
\end{equation}
We note that the restriction of the canonical $1$-form on $L$, $\Omega^1=P\cdot dX=u\cdot dq+v\cdot dp$, is
\begin{equation}
\Omega^1|_{L}=\omega^1_W=\frac{1}{2}(p\cdot dq-q\cdot dp)
\end{equation}
the Weyl-symmetrized canonical $1$-form in $\mathbb{R}^{2d}$.

\subsection{The Variational System in Double Phase Space}

The variational system is
\begin{equation}
\frac{dC}{dt}= \frac{1}{2}JH''\, C-JH''J\,D \ , \ \ \frac{dD}{dt}= -\frac{1}{4}H''\,C+ \frac{1}{2}H''J\, D
\end{equation}
with initial conditions 
\begin{equation}
C(0)=I \ , \ \ D(0)=\frac{i}{2}\,I
\end{equation}
where $H''(X):=\frac{\partial^2 H}{\partial X^2}(X)$. The argument of the Hessian element coefficients, not explicitly written, is $X_t(\alpha)$. 

The dynamics is constrained by 
\begin{equation}
\frac{d}{dt}\Big(D-\frac{1}{2}JC\Big)=0
\end{equation}
which de-couples the variational system to the non-homogeneous equations, reminiscent of the linearized Hamiltonian system 
\begin{equation}
\frac{dC}{dt}= JH''\, C-\frac{1}{2}JH''J(iI-J) \ , \ \ \frac{dD}{dt}= H''J\, D-\frac{1}{4}H''(iJ+I) \ .
\end{equation}

Just as the case in phase space $\mathbb{R}^{2d}$, the variational system is equivalent to the Riccati flow in Siegel upper half-space $\Sigma_{2d}$, by the relation $Q=DC^{-1}$,
\begin{equation}
\fl \frac{dQ}{dt}+Q\,\sigma_W(\widecheck H)_{PP}\,Q+\sigma_W(\widecheck H)_{XP}\,Q+Q\,\sigma_W(\widecheck H)_{PX}+\sigma_W(\widecheck H)_{XX}=0
\end{equation}
or, simply,
\begin{equation}
\frac{dQ}{dt}-QH''Q+\frac{1}{2}JH''Q-\frac{1}{2}QH''J+\frac{1}{4}H''=0
\label{eq:Req}
\end{equation}
with initial condition
\begin{equation}
Q(0)=\frac{i}{2}I
\end{equation}
where the arguments of the Hessian element coefficients are not explicitly written.

\subsection{Asymptotic Solution of the Canonical System in Double Phase Space}

By Maslov's complex WKB method, we construct an asymptotic solution

the asymptotic solution of the problem 
\begin{equation}
\frac{\partial F}{\partial t}+\,{}^s\! H\Big(\frac{X}{2}-J\frac{\partial F}{\partial X}\Big)=0 \ , \ \ t\geq 0
\end{equation}
and 
\begin{equation}
\frac{\partial \varphi}{\partial t}-\frac{1}{2}\,{\rm tr}\Bigg(\frac{\partial^2 F}{\partial X^2}\,{}^s\!\Big(\frac{\partial^2 H}{\partial X^2}\Big)\Big(\frac{X}{2}-J\frac{\partial F}{\partial X}\Big)\Bigg)\,\varphi=0
\end{equation}
with initial data 
\begin{equation}
F(X,Y,0)=:F_0(X,Y) \ , \ \ \varphi(Y,0;\hbar)=:\varphi_0(Y;\hbar)
\end{equation}
in the neighborhood of the zero level-set of the imaginary part of the phase. We are mostly interested in the asymptotic solution of the Weyl-symmetrized Hamilton-Jacobi equation.

Since the phase $F$ possesses a real stationary point only if $Y=g^{-t}_HX$ and $X\in\Lambda_t$, we shall be concerned only for the construction of the semi-classical asymptotic solution $\Psi^\hbar$ on $\Lambda_t$, as we expect it to be highly oscillatory, yet, exponentially small in amplitude, outside a fixed neighborhood of $\Lambda_t$, for given $t$; thus, we limit our discussion for $X$ in the neighborhood of point in $\Lambda_t$ and $Y$ in a neighborhood of $X_{-t}\in\Lambda_0$.

In the case where $Y=(\eta,\xi)\in \Lambda_0$, the asymptotic solution of the Weyl-symmetrized Hamilton-Jacobi equation is given by the following, for $(X,t)\in\Delta_Y$
\begin{eqnarray}
\fl F_{sc}(X,Y,t)=\Bigg(S_0(\eta)+F_0(X_0(\alpha),Y)+P_t(\alpha)\cdot (X-X_t(\alpha))+\int_{\Gamma^t(X_0(\alpha),P_0(\alpha))}\Omega^1\\ \nonumber
-\sigma_W(\widecheck H)(X_0(\alpha),P_0(\alpha))\,t+\frac{1}{2}(X-X_t(\alpha))\cdot Q(\alpha,t)(X-X_t(\alpha))\Bigg)_{\alpha=\alpha(X,t)}
\end{eqnarray}
where $\alpha=\alpha(X,t)$ is the unique solution of the equation 
\begin{equation}
\Big(X-X_t(\alpha)\Big)\cdot \frac{\partial X_t}{\partial \alpha_j}(\alpha)=0 \ , \ \ j=1,\ldots,d
\end{equation}
$X_t(\alpha)$ is the solution of the Characteristic System (\ref{eq:HJeq}) and (\ref{eq:TReq}) and $Q(\alpha,t)$ is the solution of the Riccati flow (\ref{eq:Req}).

The meaning of the above orthogonality relation is that for $Y\in\Lambda_0$, the neighborhood $\Delta_Y$ of $\delta_Y$ in extended phase space $\mathbb{R}^{2d+1}$ is such that for given $(X,t)\in\Delta_Y$ a unique nearest point on $\delta_Y$ is defined, with local co-ordinates $(\alpha,t)$, where $\alpha=\alpha(X,t)$ are defined by the orthogonality relation.

In particular, we consider $Y\in\Lambda_0$ and $X\in\Delta_Y$ such that 
\begin{equation}
{\rm dist}\Big((X,t),\delta_Y\Big)\leq \varepsilon(\hbar)\,\Rightarrow \, |X-X_t(\alpha)|\leq \varepsilon
\end{equation}
where the local co-ordinates $\alpha$ are defined above.

For such $(X,Y,t)$, the asymptotic solution is given by
\begin{eqnarray}
\fl F_{sc}(X,Y,t)=\Bigg(S_0(\eta)-\frac{\xi\cdot \eta}{2}+\frac{1}{2}X_0(\alpha)\cdot JY+\frac{1}{2}X\cdot J X_t(\alpha)\\ \nonumber
+\mathcal{A}(X_0(\alpha),t)-\frac{p_t(\alpha)\cdot q_t(\alpha)-p_0(\alpha)\cdot q_0(\alpha)}{2}\\ \nonumber
+\frac{i}{4}|X_0(\alpha)-Y|^2+\frac{1}{2}(X-X_t(\alpha))\cdot Q(\alpha,t)(X-X_t(\alpha))\Bigg)_{\alpha=\alpha(X,t)} 
\end{eqnarray}
while we may expand the above for the asymptoticsolution for $Y\notin\Lambda_0$, the same conditions applying for $X$.

The above is an asymptotic in the sense that it satisfied the Weyl-symmetrized Hamilton-Jacobi modulo $O({\rm Im})\,F_{sc}^{3/2}$ and the initial condition exactly, about the zero level set $\delta_Y$.

For this case the asymptotic solution of the transport equation is given simply by 
\begin{eqnarray}
\fl \varphi_{sc}(q,p,\eta,\xi,t;\hbar)=(\pi\hbar)^{-d/4}\,\frac{R_0(\eta)}{\sqrt{{\rm det}\,\Big(I-i\,S_0''(\eta)\Big)}}\\ \nonumber
\times\exp \frac{1}{2}\int\displaylimits_0^t{\rm tr}\Bigg(\frac{\partial^2 F_{sc}}{\partial X^2}(q,p,\eta,\xi,t)\,{}^s\!\Big(\frac{\partial^2 H}{\partial X^2}\Big)\Big(\frac{X}{2}-J\frac{\partial F_{sc}}{\partial X}(q,p,\eta,\xi,t)\Bigg)\Bigg)\,d\tau \ .
\end{eqnarray}

\subsection{The Asymptotic Solution of the Cauchy Problem on the Lagrangian Manifold $\Lambda_t$}

We are interested in the semi-classical asymptotic solution over the the propagated Lagrangian manifold, namely $\Psi^\hbar|_{\Lambda_t}(q,p,t)$, where 
\begin{equation}
\Lambda_t:=g^t_H\Lambda_0
\end{equation}
and $\Lambda_0$ is the Lagrangian manifold of the initial WKB data
\begin{equation}
\Lambda_0=\{(q,p)\in\mathbb{R}^{2d}|\,p=\frac{\partial S_0}{\partial q}(q)\} \ .
\end{equation}

The real stationary point equations 
\begin{equation}
{\rm Im}\,F(q,p,\eta,\xi,t)=0 \ , \ \ \frac{\partial F}{\partial \eta}(q,p,\eta,\xi,t)=0 \ , \ \ \frac{\partial F}{\partial \xi}(q,p,\eta,\xi,t)=0 
\end{equation}
have a unique solution if $(q,p)\in g^t_H\Lambda_0$, $(\eta,\xi)=g^{-t}(q,p)$. Thus, by the complex stationary phase theorem, we have 
\begin{equation}
\fl \Psi^\hbar(q,p,t)= \frac{{}^r\!\varphi(\zeta(q,p),t;\hbar)}{\sqrt{{\rm det}\,{}^r\!F''(q,p,\zeta(q,p),t)}}\,e^{\frac{i}{\hbar}\,{}^r\!F(q,p,\zeta(q,p),t)}\Big(1+o(\hbar^{1+d/4})\Big)
\end{equation}
where $\,{}^r\varphi(\zeta,t;\hbar)$ and $\,{}^rF(q,p,\zeta,t)$ are the $r$-analytic extensions to the complex variable 
\begin{equation}
\zeta=(\eta+iu,\xi+iv)\in\mathbb{C}^{2d}
\end{equation}
and $\zeta=\zeta(q,p)$ is the unique solution of the equation 
\begin{equation}
\frac{\partial\,{}^r\!F}{\partial \zeta}(q,p,\zeta,t)=0 \ .
\end{equation}
As ${\rm Im}\,F(q,p,\zeta(q,p),t)$ is everywhere non-negative, vanishing only on the Lagrangian manifold $g^t_H\Lambda_0$ when $(q,p)=g^t_H(\eta,\xi)$, the semi-classical asymptotic solution is exponentially small at fixed phase space points not belonging to $g^t_H\Lambda_0$, while along the propagated Lagrangian manifold it is 
\begin{eqnarray}
\fl \Psi^\hbar|_{\Lambda_t}(q,p,t)=(\pi\hbar)^{-d/4}\Big(\det\,\frac{\partial(\eta_t-i\xi_t)}{\partial(\eta-i\xi)}\Big)^{-1/2}_{(\eta,\xi)=g^{-t}_H(q,p)}\,R_0(q_{-t})\\ \nonumber 
\times \frac{\exp\frac{i}{\hbar}\Big(-\frac{p\cdot q}{2}+S_0(q_{-t})+\mathcal{A}(g^{-t}(q,p),t)\Big)}{\sqrt{{\rm det}\,\Big(I-iS_0''(q_{-t})\Big)\,{\rm det}\,F''(q,p,g^{-t}(q,p),t)}}\Big(1+o(\hbar)\Big) \ .
\label{eq:asympsol}
\end{eqnarray}

We note that the transported phase constrained on the real critical set, 
\begin{equation}
S(q,t)=S_0(q_{-t})+\mathcal{A}(q_{-t},p_{-t},t) 
\end{equation}
where $q_{-t}=q_{-t}(q,S'(q))$, is the solution of the corresponding \textit{Hamilton-Jacobi equation}
\begin{equation}
\frac{\partial S}{\partial t}+H\Big(q,\frac{\partial S}{\partial x}\Big)=0  
\end{equation}
with initial data 
\begin{equation}
S(x,0)=S(x)
\end{equation}
inducing the Lagrangian manifold $g^t_H\Lambda_0=:\Lambda_t$.

\section{Simple Illustrations of the Method}

In this section we provide simple illustrations of the Anisotropic Gaussian Approximation, with the construction of asymptotic solutions of the Cauchy problem for the phase space Schr\"{o}dinger equation for three different cases, which simultaneously serve as an experiment on its accuracy against the corresponding solutions. 

Physically, these examples concern simple quantum processes relevant in Atomic Physics and Quantum Optics. In particular, we consider motion in one-dimensional physical space for the case of zero potential, corresponding to free motion, the case of linear potential, corresponding to scattering off a constant electrostatic field and the case of parabolic potential, corresponding to bounded motion in a parabolic optical trap. Although these can be considered as special cases of an appropriately parametrized quadratic potential, we consider them separately, as each encapsulates different dynamical properties of a particular significance.


Sub-quadratic polynomial potentials share the common characteristic of linearity of the Hamiltonian and have identical variational system and matrix Riccati flow, yielding the --uniform in phase space-- anisotropy form 
\begin{equation}
Z(t)=\frac{i}{1+2it}
\end{equation}
while the phase space anisotropy form reads
\begin{equation}
Q(t) = \frac{i}{2(1+it)}\left(
\begin{array}{rl}
1 & -t \\
-t & 1+2it  
\end{array} \right) \ .
\end{equation}

As initial data, for all examples, we considering semi-classical WKB initial data for the Schr\"odinger equation, which allows for explicit calculations, 
\begin{equation}
\psi_0(x)=\pi^{-1/4}e^{-\frac{1}{2}x^2}e^{\frac{i}{2\hbar}x^2} 
\end{equation}
so that the initial Lagrangian manifold generated by the initial data is 
\begin{equation}
\Lambda_0=\{(q,p)\in\mathbb{R}^{2d}|\, p=q\} \ .
\end{equation}

We compare the asymptotic solutions the method produces against solutions of the phase space Schr\"odinger equation for the aforementioned simple examples, namely, the problem 
\begin{equation}
i\hbar\,\frac{\partial\Psi}{\partial t}=\Big(\frac{p}{2}-i\hbar\frac{\partial}{\partial q}\Big)^2\Psi+V\Big(\frac{q}{2}+i\hbar\,\frac{\partial }{\partial p}\Big)\Psi \ , \ \ t\in[0,T]
\end{equation}
with initial condition $\Psi(q,p,0)=\Psi_0(q,p)$, where 
\begin{equation}
\Psi_0(q,p)=\hbar^{-1/4}\sqrt{\frac{1}{\pi}\frac{1}{1-i+\hbar}}\,e^{-(q^2+ipq)/2\hbar} \exp\frac{1}{2\hbar}\frac{(q-ip)^2}{1-i+\hbar}
\end{equation}
for the zero, the linear and parabolic potentials $V$.

\subsection{Free Motion}

The Hamiltonian is
\begin{equation}
H(q,p)=p^2 
\end{equation}
which generates the Hamiltonian flow $g^t_H(q,p)=(q+2tp,p)$ , $t\in[0,T]$, while the phase space action along the Hamiltonian flow is $\mathcal{A}(q,p,t)=p^2t$.

The solution of the configuration space problem is obtained by means of the free propagator \cite{FeHi},
\begin{eqnarray}
\psi(x,t)=\frac{1}{(4\pi i \hbar t)^{1/2}}\int e^{\frac{i}{4\hbar t}(x-y)^2}\psi_0(y)\,dy 
\end{eqnarray}
which, for the particular initial data, is 
\begin{equation}
\psi(x,t)=\pi^{-1/4}\sqrt{\frac{1}{1+2(1+i\hbar)t}}\exp\frac{i}{2\hbar}\Big(\frac{1+i\hbar}{1+2(1+i\hbar)t}x^2\Big) 
\end{equation}
whose wave packet transform is 
\begin{equation}
\fl \Psi(q,p,t)=\hbar^{-1/4}\sqrt{\frac{1}{\pi}\frac{1}{1-i+\hbar(1+2it)}}\,e^{\frac{i}{2\hbar}\frac{(q-ip)\Big((i-\hbar)q-i(1+2(1+i\hbar)t)p\Big)}{1+i+2it+\hbar(i-2t)}} \ .
\label{eq:sol1}
\end{equation}

The semi-classical phase space propagator is 
\begin{eqnarray}
\fl \mathcal{K}_{sc}(q,p,\eta,\xi,t;\hbar)=\frac{1}{2\pi\hbar}\sqrt{\frac{1}{1+it}}\exp\frac{i}{\hbar}\Bigg(\frac{q\eta-p\xi}{2}-tq\xi\\ \nonumber
+\frac{i}{4(1+it)}\left(\begin{array}{ccc}
q-\eta-2t\xi\\
p-\xi 
\end{array}\right)^{{\rm T}}
\left(
\begin{array}{rl}
1 & -t \\
-t & 1+2it  
\end{array} \right)
 \left(
\begin{array}{ccc}
q-\eta-2t\xi\\
p-\xi 
\end{array}\right)\Bigg) \ .
\end{eqnarray}

It is a routine task of Gaussian integration to calculate the semi-classical asymptotic solution by means of the action of the semi-classical phase space propagator, $\mathcal{U}^t_{sc}\Psi_0$, which we unsurprisingly to be identified with the exact solution (\ref{eq:sol1}).

The semi-classical asymptotic solution --and the solution itself-- is semi-classically concentrated on the transported Lagrangian manifold by the Hamilton-Jacobi dynamics. In particular, the Hamilton-Jacobi problem 
\begin{eqnarray}
\frac{\partial S}{\partial t}+\Big(\frac{\partial S}{\partial x}\Big)^2=0 \ , \ \ t\in[0,T]\\ \nonumber
S(x,0)=S_0(x)=\frac{1}{2}x^2 
\end{eqnarray}
has the solution 
\begin{equation}
S(x,t)=\frac{1}{2}\frac{x^2}{1+2t} 
\end{equation}
which generates the transported Lagrangian manifold 
\begin{equation} 
\Lambda_t=\{(q,p)\in\mathbb{R}^{2d}|\,p=\frac{q}{1+2t}\} 
\end{equation}
i.e., initially the diagonal straight line $p=q$ asymptotically rotating clockwise to the horizontal straight line $p=0$.

\subsection{Scattering by a Constant Electrostatic Field}

The Hamiltonian is
\begin{equation}
H(q,p)=p^2+q
\end{equation}
which generates the Hamiltonian flow $g^t_H(q,p)=(q+2tp-t^2,p-t)$, $t\in[0,T]$, while the phase space action along the Hamiltonian flow is $\mathcal{A}(q,p,t)=(p^2-q)t-2pt^2+\frac{2}{3}t^3$.

The solution of the configuration space problem is obtained by means of the Airy propagator \cite{DaZh}, 
\begin{eqnarray}
\psi(x,t)=\frac{e^{-\frac{i}{\hbar}\Big(\frac{1}{3}t^3+tx\Big)}}{2(2\pi i\hbar)^{1/2}}\int e^{-\frac{1}{4i\hbar t}(x-y+t^2)}\psi_0(y)\,dy \ .
\end{eqnarray}

For the particular initial data, we obtain,
\begin{eqnarray}
\fl \psi(x,t)=\pi^{-1/4}\sqrt{\frac{1}{1+2(1+i\hbar)t}}\nonumber \\
\times \exp\frac{i}{\hbar}\Bigg(\frac{1}{2}\frac{(t^2+x)^2}{1+2(1+i\hbar)t}-\frac{1}{3}t^3-tx\Bigg) \exp-\frac{1}{2}\frac{(t^2+x)^2}{1+2(1+i\hbar)t} 
\end{eqnarray}
whose wave packet transform is 
\begin{eqnarray}
\fl \Psi(q,p,t)=\hbar^{-1/4}\sqrt{\frac{1}{\pi}\frac{1}{1-i+2t+(1+2it)\hbar}}\exp\frac{i}{6\hbar\Big(1-i+2t+(1+2it)\hbar\Big)}\\ \nonumber
\fl \times \Big(3(ip^2-(1+i)pq+\hbar pq+q^2+i\hbar q^2)-6(-ip-ip^2+\hbar p^2+q+pq+i\hbar pq)t\\ \nonumber 
\fl +3(i+2ip-2\hbar p-2q-2i\hbar q)t^2+2(-1+i-\hbar)t^3-(1+i\hbar) t^4\Big) \ .
\label{eq:sol2}
\end{eqnarray}

The semi-classical phase space propagator is 
\begin{eqnarray}
\fl \mathcal{K}_{sc}(q,p,\eta,\xi,t;\hbar)=\frac{1}{2\pi\hbar}\sqrt{\frac{1}{1+it}}\exp\frac{i}{\hbar}\Bigg(\frac{2}{3}t^3-2t^2\xi+t(\xi^2-\eta)\\ \nonumber 
\fl +\frac{i}{4(1+it)}\left(\begin{array}{ccc}
q-\eta-2t\xi+t^2\\
p-\xi+t 
\end{array}\right)^{{\rm T}}
\left(
\begin{array}{rl}
1 & -t \\
-t & 1+2it  
\end{array} \right)
\left(
\begin{array}{ccc}
q-\eta-2t\xi+t^2\\
p-\xi+t \end{array}\right)\Bigg)
\end{eqnarray}
while the semi-classical solution by the semi-classical asymptotic solution, again, is identified with the exact solution. 

The corresponding Hamilton-Jacobi problem is 
\begin{eqnarray}
\frac{\partial S}{\partial t}+\Big(\frac{\partial S}{\partial x}\Big)^2+x=0 \ , \ \ t\in[0,T]\\ \nonumber
S(x,0)=S_0(x)=\frac{1}{2}x^2 
\end{eqnarray}
has the solution 
\begin{equation}
S(x,t)=\frac{1}{2(1+2t)}\Big(x^2-2t(1+t)x-\frac{1}{3}(2+t)t^3\Big) 
\end{equation}
which generates the transported Lagrangian manifold 
\begin{equation} 
\Lambda_t=\{(q,p)\in\mathbb{R}^{2d}|\,p=\frac{q-t(t+1)}{1+2t}\} 
\end{equation}
i.e., initially the diagonal straight line $p=q$, asymptotically rotating and displacing toward the horizontal straight line $p=-\frac{t}{2}$. 

\subsection{Bound Motion in a Parabolic Optical Trap}

The Hamiltonian is
\begin{equation}
H(q,p)=p^2+q^2
\end{equation}
which generates the Hamiltonian flow $g^t_H(q,p)=R(t)(q,p)$, $t\in[0,T]$, where $R(t)=\left(
\begin{array}{rl}
\cos\,2t & \sin\,2t \\
-\sin\,2t & \cos\,2t  
\end{array} \right)$, while the phase space action along the Hamiltonian flow is $\mathcal{A}(q,p,t)=\frac{1}{4}(p^2-q^2)\sin\,4t+\frac{1}{2}pq\Big(\cos\,4t-1\Big)$.

The solution of the configuration space problem is obtained by means of the harmonic oscillator propagator \cite{FeHi}, 
\begin{equation}
\fl \psi(x,t)=\Big(\frac{1}{2\pi i \hbar\,\sin\,2t}\Big)^{1/2}\int \exp\frac{i}{2\hbar}\Big(\frac{(x^2+y^2)\,\cos\,2t-2xy}{\sin\,2t}\Big)\psi_0(y)\,dy \ .
\end{equation}

For the particular initial data, we obtain,
\begin{eqnarray}
\fl \psi(x,t)=\frac{1}{\pi^{1/4} \sqrt{\cos\,2t+(1+i\hbar)\sin\,2t}}\,\exp\frac{i}{2\hbar}\Big(\frac{-1+(1+i\hbar)\cot\,2t}{1+i\hbar+\cot\,2t}\,x^2\Big)
\end{eqnarray}
whose wave packet transform is 
\begin{eqnarray}
\fl \Psi(q,p,t)=\frac{e^{-it}}{\sqrt{\pi}\hbar^{1/4}\sqrt{1-i+\hbar}}\\ \nonumber 
\label{eq:sol3}
\times \exp\frac{i}{2\hbar}\frac{(q-ip)(-(1+i\hbar)p-q+(-p+q+i\hbar p)\cot\,2t)}{(1-i+\hbar)(i+\cot\,2t)} \ .
\end{eqnarray}

The semi-classical phase space propagator is 
\begin{eqnarray}
\fl \mathcal{K}_{sc}(q,p,\eta,\xi,t;\hbar)=\frac{1}{2\pi\hbar}e^{-it}\exp\frac{i}{\hbar}\Bigg(\frac{1}{4}(\xi^2-\eta^2)\sin\,4t+\frac{1}{2}\xi\eta(\cos\,4t-1)\\ \nonumber
+\frac{1}{2}\xi\eta(\cos^22t-\sin^22t)+\frac{1}{2}(\xi^2-\eta^2)\cos\,2t\,\sin\,2t\\ \nonumber
+\frac{1}{2}q(\xi\cos\,2t-\eta\sin\,2t)-\frac{1}{2}p(\eta\cos\,2t+\xi\sin\,2t)\\ \nonumber 
\fl+\frac{i}{4(1+it)}\left(\begin{array}{ccc}
q-\eta\cos\,2t-\xi\sin\,2t\\
p-\xi\cos\,2t+\eta\sin\,2t
\end{array}\right)^{{\rm T}}
\left(\begin{array}{rl}
1 & -t \\
-t & 1+2it  
\end{array} \right)
\left(\begin{array}{ccc}
q-\eta\cos\,2t-\xi\sin\,2t\\
p-\xi\cos\,2t+\eta\sin\,2t \end{array}\right)\Bigg)
\end{eqnarray}
while the semi-classical solution by the semi-classical asymptotic solution, again, is identified with the exact solution. 

The corresponding Hamilton-Jacobi problem is 
\begin{eqnarray}
\frac{\partial S}{\partial t}+\Big(\frac{\partial S}{\partial x}\Big)^2+x^2=0 \ , \ \ t\in[0,T]\\ \nonumber
S(x,0)=S_0(x)=\frac{1}{2}x^2 
\end{eqnarray}
has the solution 
\begin{equation}
S(x,t)=\frac{1}{2}\frac{\cos\,4t}{(\cos\,2t+\sin\,2t)^2}x^2
\end{equation}
which generates the transported Lagrangian manifold 
\begin{equation} 
\Lambda_t=\{(q,p)\in\mathbb{R}^{2d}|\,p=\frac{\cos\,4t}{(\cos\,2t+\sin\,2t)^2}q\} 
\end{equation}  
i.e., initially the diagonal straight line $p=q$, rotating clock-wise, becoming the vertical line at $t=\frac{\pi}{2}(n-\frac{1}{8})$ for $n=1,2,\ldots$ 

\section{Discussion and Conclusions}

We have developed a theory a semi-classical time evolution for the Weyl-symmetrized phase space Schr\"{o}dinger equation on the basis of a particular semi-classical approximation for the time evolution of single semi-classical isotropic Gaussian wave packets, termed the Anisotropic Gaussian Approximation. Utilizing the wave packet representation, we express the dynamics of the solution of the phase space Schr\"{o}dinger equation as a superposition of semi-classically evolved wave packets according to the above approximation. 

On this basis, we have constructed a semi-classical phase space propagator in the form of a semi-classical Fourier integral operator which induces a complex WKB canonical system in double phase space compatible with the canonical system induced on the kernel of the propagator by the canonical system of Berezin and Shubin, sharing the same characteristic Weyl symmetrization. 

Given the semi-classical phase space propagator, we show that the semi-classical asymptotic solution of the Cauchy problem for the Weyl-symmetrized phase space Schr\"{o}dinger equation is semi-classically concentrated along the transported Lagrangian manifold of the WKB initial data, along which we provide an expression for it.  

We have given a construction of a method of calculating semi-classical time evolution in phase space which is conceptually and computationally concise, which could be of interest for applications in problems of Atomic Physics, Quantum Optics and Quantum Chemistry. 

To the authors' knowledge there is no prior work on the construction of semi-classical asymptotic solutions of the phase space Schr\"{o}dinger equation for the semi-classical time evolution problem. We consider this work to be a contribution in the development of the theory, toward a deeper understanding of the phase space Schr\"{o}dinger equation and linear representations of Quantum Mechanics in phase space, linear phase space representations of other linear differential equations, such as the wave equation. 

Possible directions for a deeper understanding of the subject could be on a complete study of the asymptotic solutions of the Weyl-symmetrized Hamilton-Jacobi equation in double phase space in determining a simple expression for the asymptotic solution of the wave function along and near the propagated Lagrangian manifold. Also, an issue of theoretic interest, as well, is the structure induced by the Weyl quantization in double phase space. Finally, a rigorous treatment on the validity of the semi-classical approximation with respect to the time interval for the Cauchy problem is indeed open and necessary.

An interesting direction of further investigation would be the extension of the theory for the case of compact configuration space, which could provide a more convenient setting for numerical investigations in the special case of compact domains in $\mathbb{R}^3$, with potential applications for billiard or cavity problems in the fields of Quantum Chaos and Optics.

\section{Acknowledgments}

The authors would like to thank Sergey Dobrokhotov, Frederic Faure, Maurice de Gosson, Robert Littlejohn, Vladimir Nazaikinskii, Vesselin Petkov and Roman Schubert for fruitful conversations and enlightening comments. Also, the authors acknowledge funding of this work in its beginning by the Archimedes Center for Modeling, Analysis and Computation.

\appendix

\section{Asymptotic Inference to the van Vleck Formula}

We derive semi-classical asymptotics for the kernel of the propagator, $K$, by means of the complex stationary phase method (see \ref{CSPT}), in order to exhibit its semi-classical equivalence to the van Vleck formula (see, e.g., Robert et al. \cite{BiRo}). 

Our starting point is the semi-classical Fourier Integral representation of the semi-classical propagator,  
\begin{eqnarray}
\fl K(x,y,t;\hbar)\sim\Big(\frac{1}{2\pi\hbar}\Big)^d\int \bar G^{\hbar}_{(q,p)}(y)G^{Z,\hbar}_{(q,p)}(x,t)\,dqdp \\ \nonumber 
=\int \chi(q,p,t;\hbar)e^{\frac{i}{\hbar}\Phi(q,p,x,y,t)}\,dqdp
\end{eqnarray}
where we have the amplitude function
\begin{equation}
\chi(q,p,t;\hbar)=2^{-d}(\pi\hbar)^{-3d/2} \Big({\rm det}\,{\rm Im}\,Z(q,p,t)\Big)^{1/4}
\end{equation}
and the complex valued phase function
\begin{eqnarray}
\fl \Phi(q,p,x,y,t)=\mathcal{A}(q,p,t)-p\cdot (y-q)+\frac{i}{2}|y-q|^2\\ \nonumber
+p_t\cdot (x-q_t)+\frac{1}{2}(x-q_t)\cdot Z\,(x-q_t) \ .
\end{eqnarray} 

We proceed with a formal application of the complex stationary phase theorem, in a weak sense.

The real stationary set of the phase
\begin{equation}
\mathcal{C}_\Phi^\mathbb{R}:=\{(q,p)\in\mathbb{R}^{2d}|\,\frac{\partial \Phi}{\partial q}=0 , \, \frac{\partial \Phi}{\partial p}=0 \ \ {\rm and } \ \ {\rm Im}\,\Phi=0\}
\end{equation}
is explicitly given by the equations
\begin{eqnarray}
\fl q=y \ , \ \ q_t=x \\ \nonumber 
\fl \frac{\partial\Phi}{\partial q}=-i(y-q)+\Big(\frac{\partial p_t}{\partial q}\Big)^{\rm T}(x-q_t)-\Big(\frac{\partial q_t}{\partial q}\Big)^{\rm T} Z\,(x-q_t)+\frac{1}{2}(x-q_t)^{\rm T}\frac{\partial Z}{\partial q}(x-q_t) \\ \nonumber 
\fl \frac{\partial\Phi}{\partial p}=-(y-q)+\Big(\frac{\partial p_t}{\partial p}\Big)^{\rm T}(x-q_t)-\Big(\frac{\partial q_t}{\partial p}\Big)^{\rm T} Z\,(x-q_t)+\frac{1}{2}(x-q_t)^{\rm T}\frac{\partial Z}{\partial p}(x-q_t)
\end{eqnarray}
which, by virtue of the relations (\ref{eq:dA}) we reduce the real stationary set to the transparent form 
\begin{equation}
\mathcal{C}_\Phi^\mathbb{R}=\{(q,p)\in\mathbb{R}^{2d}|\,q=y \ \ {\rm and } \ \ q_t=x\}
\end{equation}
so that it corresponds to a countable, even finite set of orbits $\gamma$, obeying the boundary conditions $\pi\gamma(0)=y$ and $\pi\gamma(t)=x$, where $\pi:\mathbb{R}^{2d}\rightarrow \mathbb{R}^d$, is the canonical projection onto the configuration space. Equivalently, we write $(q,q_t)=(y,x)\,\Rightarrow \, x=  q(t;y,p)$.

Denoting the number of such orbits by $N_{y,x}^t:=\#\{p\in\mathbb{R}^d|\,x=  q(t;y,p)\}$, we denote the real solutions of $x=  q(t;y,p)$ for $p$, each unique for each orbit $\gamma_r$, by 
\begin{equation}
\{p_r(x,y,t)\}=\{p_1(x,y,t),p_2(x,y,t),\ldots,p_{N^t_{y,x}}(x,y,t)\}
\end{equation}
where $\gamma_r:=\gamma^t(y,p_r(x,y,t))$, for $r=1,\ldots , N^t_{y,x}$. We note the relations 
\begin{equation}
\frac{\partial p_r}{\partial x}=\frac{\partial p}{\partial q_t} \ , \ \ \frac{\partial p_r}{\partial y}=-\frac{\partial p}{\partial q_t}\frac{\partial q_t}{\partial q}
\end{equation}
which are deduced by differentiation of $x=  q(t;y,p_r(x,y,t))$.

It is straightforward to incur the Hessian matrix of the phase $\Phi$, with respect to $(q,p)$, on the real critical set $\mathcal{C}_\Phi^\mathbb{R}$,
\begin{equation}
\fl \Phi''=\left(
\begin{array}{ccc}
iI-\Big(\frac{\partial p_t}{\partial q}\Big)^{\rm T}\frac{\partial q_t}{\partial q}+\Big(\frac{\partial q_t}{\partial q}\Big)^{\rm T} Z\,\frac{\partial q_t}{\partial q} & -\Big(\frac{\partial p_t}{\partial q}\Big)^{\rm T}\frac{\partial q_t}{\partial p}+\Big(\frac{\partial q_t}{\partial q}\Big)^{\rm T} Z\,\frac{\partial q_t}{\partial p}  \\
-\Big(\frac{\partial q_t}{\partial p}\Big)^{\rm T}\frac{\partial p_t}{\partial q}+\Big(\frac{\partial q_t}{\partial p}\Big)^{\rm T} Z\,\frac{\partial q_t}{\partial q} & -\Big(\frac{\partial p_t}{\partial p}\Big)^{\rm T}\frac{\partial q_t}{\partial p}+\Big(\frac{\partial q_t}{\partial p}\Big)^{\rm T} Z\,\frac{\partial q_t}{\partial p}
\end{array}\right) \ .
\end{equation}

The determinant of the Hessian matrix over the real stationary set is 
\begin{equation}
{\rm det}_{\mathbb{C}^{2d\times 2d}}\Phi''={\rm det}_{\mathbb{C}^{d\times d}}\Bigg(\frac{2}{i}\Big(\frac{\partial q_t}{\partial q}+i\frac{\partial q_t}{\partial p}\Big)^{-1}\frac{\partial q_t}{\partial p}\Bigg)_{(q,q_t)=(y,x)}
\end{equation}
which is non-singular, as $\frac{\partial q_t}{\partial q}+i\frac{\partial q_t}{\partial p}$ is non-singular everywhere in phase space, for any $t\geq 0$, while $\frac{\partial q_t}{\partial p}$ is non-singular for $0\leq t\leq T$, assuming that no caustics or focal point develop within this time interval.

By applying the real stationary phase lemma 
\begin{equation}
{\rm Im}\,Z=\Bigg(\frac{\partial q_t}{\partial q}\Big(\frac{\partial q_t}{\partial q}\Big)^{\rm T}+\frac{\partial q_t}{\partial p}\Big(\frac{\partial q_t}{\partial p}\Big)^{\rm T}\Bigg)^{-1}
\end{equation}
we have 
\begin{equation}
\fl K(x,y,t;\hbar)\sim\Big(\frac{1}{2\pi i\hbar}\Big)^{d/2}\sum_{r=1}^{N_{y,x}^t}\sqrt{\Big|\det\,\frac{\partial p}{\partial q_t}(t;y,p_r(x,y,t))\Big|}\,e^{\frac{i}{\hbar}\mathcal{A}(y,p_r(x,y,t),t)-\frac{\pi i }{2}\nu_r}
\end{equation}
semi-classically, in an appropriate weak sense.

In the above, the index $r=1,\ldots,N_{y,x}^t$ indexes all trajectories emanating from point $x$ at time $0$ and terminating at point $y$ at time $t$, while 
\begin{equation}
\nu_r={\rm ind}\Bigg(\Big(\frac{\partial q_t}{\partial q}+i\frac{\partial q_t}{\partial p}\Big)^{-1}\frac{\partial p}{\partial q_t}\Bigg):=\sum_{\lambda}{\rm Arg}(\lambda)
\end{equation}
for $\lambda\in\sigma\Big((\frac{\partial q_t}{\partial q}+i\frac{\partial q_t}{\partial p})^{-1}\frac{\partial p}{\partial q_t}\Big)$, i.e., the index of the monodromy matrix, or van Vleck matrix, the excess of its positive over negative eigenvalues.

Alternatively, one may give a different expression of the van Vleck matrix, as 
\begin{equation}
\frac{\partial p}{\partial q_t}(t;y,p_r(x,y,t))=\frac{\partial^2}{\partial x\partial y}\mathcal{A}(y,p_r(x,y,t),t)
\end{equation}
so that 
\begin{equation}
\fl K(x,y,t;\hbar)\sim\Big(\frac{1}{2\pi i\hbar}\Big)^{d/2}\sum_{r=1}^{N_{y,x}^t}\sqrt{\Big|\det\,\frac{\partial^2}{\partial x\partial y}\mathcal{A}(y,p_r(x,y,t),t)\Big|}\,e^{\frac{i}{\hbar}\mathcal{A}(y,p_r(x,y,t),t)-\frac{\pi i }{2}\nu_r}
\end{equation}
which is the more usual form of the semi-classical van Vleck formula. 

\section{Construction of the Wave Packet Transform}
\label{WPTr}

The Weyl shifts, $\mathcal{T}^\hbar_{(q,p)}$, define a non-commutative unitary group \cite{Fol,Lit2,CoRo} with group composition property, for $(q,p),(\eta,\xi)\in\mathbb{R}^{2d}$,
\begin{equation}
\mathcal{T}^\hbar_{(q,p)}\mathcal{T}^\hbar_{(\eta,\xi)}=e^{\frac{i}{2\hbar}(q,p)\cdot J(\eta,\xi)}\mathcal{T}^\hbar_{(q+\eta,p+\xi)}
\end{equation}
which acts on $L^2(\mathbb{R}^d)$ as
\begin{equation}
\mathcal{T}_{(q,p)}^\hbar\psi(x)=\exp\frac{i}{\hbar}\Big(p\cdot (x-q)+\frac{p\cdot q}{2}\Big)\psi(x-q)
\end{equation}
where the phase $\frac{1}{2}p\cdot q$ is due to the non-commutativity of the Heisenberg algebra.

The Weyl are the `building blocks' for the construction of the set of semi-classical isotropic Gaussian wave packets. Initially, we define the isotropic Gaussian vacuum state as the normalized state which saturates the Heisenberg inequalities as follows 
\begin{equation}
\Delta_{G_0^\hbar} \widehat q_j=\Delta_{G_0^\hbar} \widehat p_j=\sqrt{\frac{\hbar}{2}}
\end{equation}
for $j=1,\ldots, d,$ with expectations $\langle G_0^\hbar,\widehat q\,G_0^\hbar\rangle=0$ and $\langle G_0^\hbar,\widehat p\,G_0^\hbar\rangle=0$,
\begin{equation}
G_0^\hbar(x):=(\pi\hbar)^{-d/4}e^{-|x|^2/2\hbar} \ .
\end{equation} 

Subsequently, we construct the family of semi-classical isotropic Gaussian wave packets $\{G_{(q,p)}^\hbar\}$ by the action
\begin{equation}
G^\hbar_{(q,p)}(x):=\mathcal{T}^\hbar_{(q,p)}G^\hbar_0(x)
\end{equation}
where 
\begin{equation}
G^\hbar_{(q,p)}(x)=(\pi\hbar)^{-d/4}\exp\frac{i}{\hbar}\Big(\frac{p\cdot q}{2}+p\cdot (x-q)+\frac{i}{2}|x-q|^2\Big) \ .
\end{equation}
In analogy to the Gaussian vacuum state, they are normalized and minimal uncertainty states, in the sense that $\Delta_{G_{(q,p)}^\hbar}\widehat q_j\,\Delta_{G_{(q,p)}^\hbar}\widehat p_j=\frac{\hbar}{2}$.

There exists is a multitude of definitions of semi-classical wave packets, which are equivalent modulo phase $\phi(q,p)$, centered \cite{Fol,Lit2} at the origin $\phi(0)=0$. In our convention $\phi(q,p)=\frac{1}{2}p\cdot q$.

The totality of semi-classical anisotropic Gaussian wave packets $\{G^\hbar_{(q,p)}\}$ constitute a Parseval Heisenberg-Weyl frame \cite{Gro,TAG}, an over-complete set in $L^2(\mathbb{R}^d)$. They define the wave packet transform, 
\begin{equation}
\mathcal{W}\psi(q,p)=\Big(\frac{1}{2\pi\hbar}\Big)^{d/2}\int\bar G_{(q,p)}^\hbar(x)\psi(x)\,dx
\end{equation}
which is an isometric isomorphism $L^2(\mathbb{R}^d)\rightarrow \mathfrak{F}$, in the sense that the Plancherel theorem is satisfied, 
\begin{equation}
\int|\mathcal{W} \psi(q,p)|^2\,dqdp=\int|\psi(x)|^2\,dx=1
\end{equation}
and the Parseval formula
\begin{equation}
\int\bar \psi\varphi  \,dx=\int\overline{\mathcal{W}\psi}\, \mathcal{W}\varphi  \, dqdp \ .
\end{equation}

Semi-Classical isotropic Gaussian wave packets constitute a non-denumerable over-complete set in $L^2(\mathbb{R}^d)$, in the sense that any $\psi\in L^2(\mathbb{R}^d)$ can be expressed uniquely as a superposition of coherent states, so that
\begin{equation}
\Big(\frac{1}{2\pi\hbar}\Big)^d\int G_{(q,p)}^\hbar\langle G_{(q,p)}^\hbar,\cdot\rangle \,dqdp={\rm Id}
\end{equation}
which is a completeness relation in $L^2(\mathbb{R}^{2d})$ \cite{NaSt,Rob1}, 
\begin{equation}
\bigg(\frac{1}{2\pi\hbar}\bigg)^{d} \int\bar G_{(q,p)}^\hbar(x)G_{(q,p)}^\hbar(y)\,dqdp=\delta(x-y) \ .
\end{equation}
They are, however, not orthogonal in $L^2(\mathbb{R}^d)$ \cite{NSS,NaSt,Rob1},
\begin{equation}
\langle G_{(q,p)}^\hbar,G_{(\eta,\xi)}^\hbar\rangle=\exp\frac{i}{\hbar}\Bigg(\frac{1}{2}(q,p)\cdot J(\eta,\xi)+\frac{i}{4}\Big(|q-\eta|^2+|p-\xi|^2\Big)\Bigg)
\end{equation}
but, rather, they are asymptotically orthogonal
\begin{equation}
\bigg(\frac{1}{2\pi\hbar}\bigg)^d\langle G_{(q,p)}^\hbar,G_{(\eta,\xi)}^\hbar\rangle\rightharpoonup_*\delta(q-\eta)  \delta(p-\xi) \ , \ \ \hbar\rightarrow 0^+ \ .
\end{equation} 

The adjoint operator of $\mathcal{W}$ \cite{NaSt} is defined as 
\begin{eqnarray}
\mathcal{W}^*\Phi(x)=\bigg(\frac{1}{2\pi\hbar}\bigg)^{d/2}\lim_{n\rightarrow \infty}\int G_{(q,p)}^\hbar(x)\Phi_n(q,p)\,dqdp \\ \nonumber
=:\bigg(\frac{1}{2\pi\hbar}\bigg)^{d/2}\int^*G_{(q,p)}^\hbar(x)\Phi(q,p)\,dqdp 
\label{eq:star}
\end{eqnarray}
where $\{\Phi_n\}_{n\in\mathbb{N}}$ is a sequence of bounded, compactly supported phase space functions with $\Phi_n\rightarrow \Phi$, strongly in $L^2(\mathbb{R}^{2d})$.
The adjoint enables us to introduce the orthogonal projection $\mathcal{P}_{\mathfrak{F}}:L^2(\mathbb{R}^{2d})\rightarrow\mathfrak{F}$ \cite{NaSt},
\begin{equation}
\mathcal{P}_{\mathfrak{F}}:=\mathcal{W} \mathcal{W} ^* \ .
\label{eq:projector1}
\end{equation} 
Restricted on $\mathfrak{F}$, the adjoint $\mathcal{W}^*$ is identified with the inverse operator $\mathcal{W}^{-1}$.

The wave packet transform micro-localizes quantum states in phase space; the real part of the phase of the kernel contributes to semi-classical localization in the momenta, while the imaginary part, to semi-classical localization in the positions.

The initial phase space propagator is the kernel of the projector $\mathcal{P}_{\mathfrak{F}}$, meaning that if $\Psi\in\mathfrak{F}$, then
\begin{equation}
\fl \mathcal{U}^0\Psi(q,p)=\int\mathcal{K}(q,p,\eta,\xi,0;\hbar)\Psi (\eta,\xi)\,d\eta d\xi=\mathcal{P}_{\mathfrak{F}}\Psi (q,p)=\Psi (q,p) 
\end{equation}
so that for $t=0$ the representation of the solution $\Psi (q,p,t)$ reconstructs the initial data, $\Psi_0$, 
\begin{eqnarray}
\fl \Psi (q,p,0)=\Big(\frac{1}{2\pi\hbar}\Big)^{d} \int \!\! \int\bar G_{(q,p)}^\hbar(x)G _{(\eta,\xi)}^\hbar(x)\Psi_0 (\eta,\xi)\,dxd\eta d\xi \\ \nonumber
\fl =\Big(\frac{1}{2\pi\hbar}\Big)^{3d/2}\int\psi_0 (y)\Big(\int \!\! \int\bar G_{(q,p)}^\hbar(x)G _{(\eta,\xi)}^\hbar(x)\bar G _{(\eta,\xi)}(y;\hbar)\, dxd\eta d\xi\Big)\,dy \\ \nonumber
=\Big(\frac{1}{2\pi\hbar}\Big)^{d/2}\int\bar G _{(q,p)}^\hbar(x)\psi_0 (x)\,dx=\Psi _0(q,p)  
\end{eqnarray}
by utilizing the completeness property. The kernel, $b$, known as the Bergmann reproducing kernel, is the kernel of the projector $\mathcal{P}_{\mathfrak{F}}$ restricted on $\mathfrak{F}$, and assumes the integral form 
\begin{equation}
b(q,p,\eta,\xi;\hbar)=\Big(\frac{1}{2\pi\hbar}\Big)^{d}\int \bar G^\hbar_{(q,p)}(x)G^\hbar_{(\eta,\xi)}(x)\,dx 
\end{equation}
noting that $b(q,p,\eta,\xi;\hbar)=\mathcal{K}_{sc}(q,p,\eta,\xi,0;\hbar)$, with the defining property 
\begin{equation}
\int b(q,p,\eta,\xi;\hbar)\Psi(\eta,\xi)\,d\eta d\xi=\Psi(q,p) 
\end{equation}
for $\Psi\in\mathfrak{F}$.

The local properties of $\mathcal{P}_{\mathfrak{F}}$ are a `smoothing' on the Heisenberg scale, smearing away oscillations on that scale. For $\Phi\in L^2(\mathbb{R}^{2d})$
\begin{equation}
\mathcal{P}_{\mathfrak{F}}\Psi(q,p)=\Big(\frac{1}{2\pi\hbar}\Big)^d\int \bar G^\hbar_{(q,p)}(x)\Big(\int^* G^\hbar_{(\eta,\xi)}(x)\Phi (\eta,\xi)\,d\eta d\xi\Big)\,dx \ .
\label{eq:Berg1}
\end{equation}

The wave packet $G_{(q,p)}^\hbar$ has the expectations
\begin{equation}
\langle G_{(q,p)}^\hbar, \widehat q\,G_{(q,p)}^\hbar\rangle=q \ , \ \ \langle G_{(q,p)}^\hbar,\widehat p\,G_{(q,p)}^\hbar\rangle=p \ .
\end{equation}
Its physical significance lies in that it is the unique optimal wave function minimizing the product of the position-momentum uncertainties in a uniform way for all positions and momenta, a property which naturally associates it with a pure classical state, i.e., a point in phase space.

Even though $|\Psi(q,p)|^2\,dqdp$ possesses the characteristics of a phase space probability density, it does not allow for a statistical interpretation of $\Psi$ analogous to that of the Schr\"{o}dinger wave function, i.e., a probability amplitude of a particle's simultaneous position and momentum being $(q,p)\in\mathbb{R}^{2d}$, as the maximum quantum resolution in phase space is the Planck cell. The physical content of the phase space Schr\"{o}dinger wave function, $\Psi$, is that it is the objective description of the particle in question, representing the potential possibility of the particle \textit{being} in the state $G_{(q,p)}^\hbar$ \cite{Fok}.

As for the Fock-Bargmann decomposition of the phase space Hilbert space, there is, further, a simple physical interpretation. In particular, the decomposition
\begin{equation}
L^2(\mathbb{R}^{2d})=\mathfrak{F}\oplus \mathfrak{F}^{\perp} 
\end{equation} 
interpreted as a decomposition of of square integrable phase space functions which are images of Schr\"{o}dinger wave functions under the wave packet transform and those which are not. 

As only those $\Psi\in\mathfrak{F}$ correspond to Schr\"{o}dinger wave functions on configuration space, only `twisted' analytic phase space functions correspond to \textit{pure quantum states}, while all other choices of square integrable phase space functions correspond to \textit{mixed quantum states}.

The wave packet transform is related to the \textit{Bargmann transform} \cite{Bar1,Bar2,NSS,NaSt} by
\begin{equation}
\mathcal{W}\psi(q,p)=(2\pi\hbar^{1/2})^{-d/2}\exp\Big(-\frac{i}{\hbar}\frac{p\cdot q-i|p|^2}{2}\Big)\Big(e^{-z^2}\mathcal{B}f(z)\Big)_{z=\frac{q-ip}{\sqrt{2\hbar}}}
\end{equation}
where $\mathcal{B}f$ is the Bargmann transform of $f(x)=\psi(\sqrt{\hbar}x)$, 
\begin{equation}
\mathcal{B}f(z)=\frac{1}{\pi^{d/4}}\int e^{-\frac{1}{2}(z^2+|x|^2)+\sqrt{2}z\cdot x}f(x)\,dx \ .
\end{equation}


\section{Quantizations and Representations}
\label{QuRep}

Although semi-classical asymptotic solutions of the Schr\"{o}dinger equation all share a common semi-classical limit, in some sense, substantial differences may arise between sub-leading asymptotic terms of the same order, in some sense, with respect to choice of quantization. Such differences become apparent in treatments of semi-classical wave packet dynamics, as are the differences in form between the semi-classically propagated wave packets considered by Robert in \cite{Rob1}, who has assumed the Weyl quantization, and those considered by Nazaikinskii et al. in \cite{NSS}, who have assumed the normal quantization. 

The most common choice of quantization in the literature, mainly on the basis of practical reasons, is the \textit{normal quantization,} for which the formal rule of quantizing physical quantities corresponding to symbols of appropriate smoothness and growth properties, $f\mapsto {\rm Op}_N(f)$, is \cite{BeSh,KaMa,Mas1,MaNa}
\begin{equation}{\rm Op}_N(f)\psi(x):=\Big(\frac{1}{2\pi\hbar}\Big)^{d}\int \!\! \int e^{\frac{i}{\hbar}p\cdot (x-q)}f(x,p)\psi(q)\,dqdp\end{equation}
or, equivalently \cite{KaMa}
\begin{equation}
{\rm Op}_N(f)=F_\hbar^{*} f F_\hbar=f\Big(\stackrel{2}{x},-i\hbar\, \stackrel{1}{\frac{\partial}{\partial x}}\Big)
\end{equation}
where $F_\hbar$ is the semi-classical Fourier transform, $F_\hbar f(p):=\Big(\frac{1}{2\pi\hbar}\Big)^{d/2}\int e^{-\frac{i}{\hbar}p\cdot x}f(x)\,dx$. For real analytic physical quantities, we have the Taylor series representation
\begin{equation}
f\Big(\stackrel{2}{x},-i\hbar\,\stackrel{1}{\frac{\partial }{\partial x}}\Big)=\sum_{\alpha,\beta\in\mathbb{N}^d_0}\frac{\partial^{\alpha}_q\partial^{\beta}_{p}f(0)}{\alpha!\beta!}x^{\alpha}\Big(-i\hbar\,\frac{\partial }{\partial x}\Big)^{\beta} \ .
\end{equation}

For the \textit{Weyl quantization}\footnote{One can define a 1-parameter family of quantizations, $f\mapsto {\rm Op} _\lambda(f)$ for $\lambda\in[0,1]$, as \cite{Gos2}
\begin{equation*}
{\rm Op} _\lambda(f)\psi(x)=\Big(\frac{1}{2\pi\hbar}\Big)^{d}\int \!\! \int e^{\frac{i}{\hbar}p\cdot (x-q)}f\Big(\lambda x+(1-\lambda)q,\xi\Big)\psi(q)\,dqdp
\end{equation*}
where $\lambda=0$ and $\lambda=1$ correspond to the normal and anti-normal quantizations, while the intermediate value $\lambda=\frac{1}{2}$ corresponds to the Weyl quantization.}, we have $f\mapsto \widehat f:={\rm Op} _{ W}(f)$ where 
\begin{equation}
\widehat f\psi(x):=\Big(\frac{1}{2\pi\hbar}\Big)^{d}\int \!\! \int e^{\frac{i}{\hbar}p\cdot (x-q)}f\Big(\frac{x+q}{2},p\Big)\psi(q)\,dqdp
\end{equation}
or, equivalently \cite{KaMa} 
\begin{equation}
\widehat f=f\Big(\stackrel{\boldsymbol{\omega}}{x},-i\hbar\, \stackrel{\boldsymbol{\omega}}{\frac{\partial}{\partial x}}\Big):=f\Big(\frac{\stackrel{1}{x}+\stackrel{3}{x}}{2},-i\hbar\, \stackrel{2}{\frac{\partial }{\partial x}}\Big) \ .
\end{equation}
For real analytic physical quantities, we have the Taylor series representation
\begin{equation}
f\Big(\stackrel{\boldsymbol{\omega}}{x},-i\hbar\, \stackrel{\boldsymbol{\omega}}{\frac{\partial}{\partial x}}\Big)=\sum_{\alpha,\beta\in\mathbb{N}^d_0}\frac{\partial^{\alpha}_q\partial^{\beta}_{p}f(0)}{\alpha!\beta!}\Big[x^{\alpha}\Big(-i\hbar\,\frac{\partial }{\partial x}\Big)^{\beta}\Big]_W 
\end{equation}
$[\cdot ]_W$ being the Weyl operator ordering of its polynomial argument
\begin{equation}
\Big[x^\alpha\Big(-i\hbar\, \frac{\partial }{\partial x}\Big)^\beta\Big]_W:=\frac{1}{2^{|\alpha|}}\sum_{\gamma\leq \alpha}\frac{\alpha!}{\gamma!(\alpha-\gamma)!}x^\gamma\Big(-i\hbar\, \frac{\partial }{\partial x}\Big)^\beta x^{\alpha-\gamma}
\end{equation} 
yielding 
\begin{equation}
\fl f\Big(\frac{\stackrel{1}{x}+\stackrel{3}{x}}{2},-i\hbar\, \stackrel{2}{\frac{\partial}{\partial x}}\Big) =\sum_{\alpha,\beta,\gamma\in\mathbb{N}_0^d|\,\gamma\leq \alpha}\frac{\partial _q^\alpha\partial _p^\beta H(0)}{2^{|\alpha|}\beta!\gamma!(\alpha-\gamma)!}x^\gamma\Big(-i\hbar\, \frac{\partial }{\partial x}\Big)^\beta x^{\alpha-\gamma} \ .
\end{equation}

A prominent difference between the normal, or other intermediate quantizations, and the Weyl quantization is the property of self-adjointness of the quantized physical quantities, such as the Hamiltonian, a property which guarantees the unitarity of the Schr\"{o}dinger flow, a property of time evolution we consider crucial in the frame of the semi-classical time evolution problem, and stands as a sufficient criterion for the choice of the Weyl quantization. In this work we choose the Weyl quantization.

The \textit{wave packet quantization} of $f$ is defined according to the rule \cite{NaSt}
\begin{equation}
{\rm Op}_{wp}(f):=\mathcal{W}^* f \mathcal{W}=\mathcal{W}^{-1} \mathcal{P}_{\mathfrak{F}} f \mathcal{W}
\end{equation}
while the integral representation is given by 
\begin{equation}
\fl {\rm Op}_{wp}(f)\psi(x)=\Big(\frac{1}{2\pi\hbar}\Big)^d\int\!\!\int^*G_{(q,p)}^\hbar(x)\bar G_{(q,p)}^\hbar(y)f(q,p)\psi(y)\,dydq dp 
\end{equation}
according to definition $(\ref{eq:star})$.

The quantizations of the canonical pairs $q_j$ and $p_j$ are all identified 
\begin{equation}
\widehat q_j=x_j \ , \ \ \widehat p_j=-i\hbar\,\frac{\partial}{\partial x_j} \ , \ \ j=1,\ldots,d
\end{equation}
the first being the multiplication operator by $x_j\in\mathbb{R}$.

All operators over some subspace of $L^2(\mathbb{R}^d)$ can be expressed in phase space, by means of the \textit{wave packet representation,} conjugating with the wave packet transform,
\begin{equation}
A\mapsto \mathcal{W}A\mathcal{W}^{-1}
\end{equation}
where we name $\mathcal{W}A\mathcal{W}^{-1}$ the wave packet representation of $A$. In particular, for a symbol $f$
\begin{equation}
\widehat f\mapsto \widecheck f:=\mathcal{W}\widehat f\mathcal{W}^{-1} \ .
\end{equation}

For the canonical pair $q$ and $p$, we have the phase space representation 
\begin{eqnarray}
\mathcal{W} x \mathcal{W}^{-1}=\frac{q}{2}+i\hbar\,\frac{\partial}{\partial p} \\ \nonumber 
\mathcal{W}\Big(-i\hbar\,\frac{\partial }{\partial x}\Big) \mathcal{W}^{-1}=\frac{p}{2}-i\hbar\,\frac{\partial}{\partial q}
\end{eqnarray}
the \textit{Bopp shifts} \cite{Bop}, characteristic of the specific choice of the wave packet transform, i.e., the particular construction of the semi-classical isotropic Gaussian wave packets by means of the Weyl shift.

In particular, the wave packet representation of the Weyl quantization of a physical quantity $f$ is given by 
\begin{equation}
\mathcal{W}f\Big(\stackrel{\boldsymbol{\omega}}{x},-i\hbar\,\stackrel{\boldsymbol{\omega}}{\frac{\partial}{\partial x}}\Big)\mathcal{W}^{-1}=f\Big(\stackrel{\boldsymbol{\omega}}{\frac{q}{2}}+i\hbar \,\stackrel{\boldsymbol{\omega}}{\frac{\partial}{\partial p}},\stackrel{\boldsymbol{\omega}}{\frac{p}{2}}-i\hbar \,\stackrel{\boldsymbol{\omega}}{\frac{\partial}{\partial q}}\Big) \ .
\end{equation}

Actually, the wave packet representation of the Weyl quantization of $f(q,p)$ yields the Weyl quantization in \textit{double phase space} of 
\begin{equation}
f\Big(\frac{q}{2}-v,\frac{p}{2}+u\Big) \ .
\end{equation}
Compactly expressed, for $X=(q,p)$ and $P=(u,v)$
\begin{equation}
\widecheck f\Psi(X)=\Big(\frac{1}{2\pi\hbar}\Big)^{2d}\int e^{\frac{i}{\hbar}P\cdot (X-Y)}\sigma_W(\widecheck f)\Big(\frac{X+Y}{2},P\Big)\Psi(Y)\, dYdP
\end{equation}
where the symbol is 
\begin{equation}
\sigma_{{\rm W}}(\widecheck f)(X,P)=f\Big(\frac{X}{2}-JP\Big)=f\Big(\frac{q}{2}-v,\frac{p}{2}+u\Big) \ .
\end{equation}

An important result is that the wave packet representations of the normal and Weyl orderings bare the same symbols and operators in double phase space
\begin{equation}
\sigma_W(\widecheck H)=\sigma_N(\widecheck H) , \ \ \widecheck H={\rm Op}_W(\sigma_W(\widecheck H))={\rm Op}_N(\sigma_N(\widecheck H))
\end{equation}
meaning that 
\begin{equation}
f\Big(\stackrel{\boldsymbol{\omega}}{\frac{X}{2}}+i\hbar J\stackrel{\boldsymbol{\omega}}{\frac{\partial}{\partial X}}\Big)=f\Big(\stackrel{2}{\frac{X}{2}}+i\hbar J\stackrel{1}{\frac{\partial}{\partial X}}\Big) \ . 
\end{equation}

\section{Relation to Alternative Phase Space Formulations}
\label{wignr}

As was noted by Torres-Vega et al. \cite{TZM}, Chruscinski and Mlodawksi \cite{ChMl} and later established by de Gosson \cite{Gos3}, there is a multitude of equivalent linear representations of Quantum Mechanics in phase space, leading to representations of the equations which constitute the family of phase space Schr\"{o}dinger equations, is realized by the \textit{windowed wave packet transform}, 
\begin{equation}
\mathcal{W}_\varphi \psi(q,p)=\int\varphi^\hbar _{(q,p)}(x)\psi(x)\, dx 
\end{equation}
where the `window' function, $\varphi _{(q,p)}^\hbar\in\mathcal{S}(\mathbb{R}^d)$ with $\|\varphi _{(q,p)}^\hbar\|=1$, is mirolocalized at $(q,p)\in\mathbb{R}^{2d}$ on the Heisenberg scale. Different choices of window function correspond to different operators and thus different representations of the Schr\"{o}dinger equation in phase space, different formal rules of quantization in phase space. This formulation includes the wave packet formulation, for a specific Gaussian choice of window function. 

The above transformation is widely used in the area of Harmonic Analysis termed Time-Frequency Analysis (see, e.g., \cite{Gro}).

The windowed wave packet representation, however, including the wave packet representation we consider, cannot be used for the description of microscopic systems in strong interaction with their external environment. In this case, a wave function description is not possible; the system is necessarily described by a \textit{statistical ensemble} of wave functions, leading to the notion of a \textit{Blokhintsev density operator} \cite{Blo,Fok,Wig,ZFC}, a trace class rank one projection operator of the form
\begin{equation}
\rho =\sum_{\psi\in\mathcal{E}}m(\psi) \mathcal{P}_{\psi} 
\end{equation}
over some countable wave function ensemble, $\mathcal{E}$, where $\mathcal{P}_{\psi}$ is the orthogonal projection onto the ray ${\rm span}\{\psi\}$, and $m(\psi)$ are probabilities weights. In the case of an isolated quantum system, the density operator reduces to $\rho=\psi\langle \psi,\cdot \rangle$.

The Wigner function is defined by the Wigner transform of the wave function \cite{ZFC},
\begin{eqnarray}
W(q,p;\hbar)=\bigg(\frac{1}{2\pi\hbar}\bigg)^{d}\int e^{\frac{i}{\hbar}p\cdot x}\psi \Big(q-\frac{x}{2}\Big)\bar \psi \Big(q+\frac{x}{2}\Big)\,dx \\ \nonumber
\fl =\bigg(\frac{1}{2\pi\hbar}\bigg)^{d}\int W(G_{(\eta,\xi)}^\hbar,G_{(u,v)}^\hbar)(q,p;\hbar)\Psi (\eta,\xi,t)\bar\Psi (u,v,t)\,d\eta d\xi du dv 
\label{eq:wignerps}
\end{eqnarray}
where 
\begin{equation}
\fl W(G_{(\eta,\xi)}^\hbar,G_{(u,v)}^\hbar)(q,p)=\Big(\frac{1}{2\pi\hbar}\Big)^{d}\int e^{\frac{i}{\hbar}p\cdot x}G_{(\eta,\xi)}^\hbar\Big(q-\frac{x}{2}\Big)\bar G_{(u,v)}^\hbar\Big(q+\frac{x}{2}\Big)\,dx \ .
\end{equation}

Unlike the wave packet transform, the Wigner transform defines a \textit{bi-linear} integral transform of the wave function. It is also real valued, unlike the complex valued phase space wave function.

There is a direct connection between the Wigner formulation and wave packet quantization. Although it is appealing, the phase space density $|\Psi|^2$ is not identified with the Wigner function, which assumes both positive and negative values. The relation is given through the corresponding \textit{Husimi density} \cite{ZFC},
\begin{equation}
h(q,p)=|\Psi (q,p)|^2=g * W(q,p) 
\end{equation}
a the convolution of the Wigner function with the Gaussian 
\begin{equation}
g(q,p;\hbar)=(\pi\hbar)^{-d}\exp\Big(-\frac{|q|^2+|p|^2}{\hbar}\Big) 
\end{equation} 
localized over the Heisenberg scale, smearing away oscillations on that scale which `sweep' both positive and negative signs, rendering it a true phase space density.


\section{Asymptotics of Integrals of Rapidly Oscillating Functions}
\label{CSPT}

In the present, we give the fundamental result on the asymptotic behavior of highly oscillating integrals with complex phase function, of the form resulting from the construction of WKB wave functions, a version of the method of stationary phase for complex phase functions. In particular, we introduce the notion of $s$-analytic extensions and the complex stationary phase theorem, following the works of Fedoriuk \cite{Fed1,Fed2}, Le Vu \cite{LeV}, Nazaikinskii et al. \cite{NOSS} and Mishchenko et al. \cite{MSS}. Sj\"{o}strand and Melin \cite{MeSj} reach the same result along somewhat different lines. 

Let $f\in C^r(\mathbb{R}^n,\mathbb{C})$ and $s\in\mathbb{N}_0$ for $s\leq r$. We define its \textit{$s$-analytic extension}, $^{s}\! f:\mathbb{C}^n\rightarrow \mathbb{C}$, as follows \cite{Mas1,MSS,NOSS},
\begin{equation}
{}^s \! f(x+iy):=\sum_{m=0}^s\frac{1}{m!}\left(iy\cdot \frac{\partial}{\partial x}\right)^mf(x) = \sum_{\alpha\in\mathbb{N}_0^n}\sum_{|\alpha|=0}^s\frac{1}{\alpha!}(iy)^\alpha\partial_x^\alpha f(x)
\end{equation}
for $s>0$, while 
\begin{equation}
{}^0 \! f(x+iy):=f(x) 
\end{equation}
in the complex domain
\begin{equation}
\mathcal{O}:=\{z=x+iy\in\mathbb{C}^n|\,x,y\in\mathbb{R}^n\} \ .\end{equation}

For the derivatives, the following relations hold for $1\leq s\leq r-1$, 
\begin{equation}\frac{\partial\,{}^s \! f}{\partial x}(z)=\,{}^s \!\Big(\frac{\partial f}{\partial x}\Big)(z)\end{equation}
and
\begin{equation}\frac{\partial\,{}^s \! f}{\partial y}(z)=i\,{}^s \!\Big(\frac{\partial f}{\partial x}\Big)(z)-\frac{i}{s!}\Big(iy\cdot \frac{\partial }{\partial x}\Big)^s\,\frac{\partial f}{\partial x}(x) \ .
\end{equation}
It is straightforward to incur that
\begin{equation}\frac{\partial\,{}^s \! f}{\partial \bar z}(z)=O(|y|^s) \ , \ \ |y|\rightarrow 0^+
\end{equation}
point-wise in $\mathcal{O}$. 
\\

Before we proceed to the main result on the asymptotic behavior of oscillating integrals, we note the case of Gaussian integrals. For non-singular $M\in\mathbb{C}^{m\times m}$ with $M^{{\rm T}}=M$, ${\rm Re } \,M\succ 0$ and $v\in\mathbb{C}^m$, we can give a closed form expression for the Gaussian integral, where $\hbar>0$,
\begin{equation} 
\fl \bigg(\frac{1}{2\pi\hbar}\bigg)^{m/2}\int_{\mathbb{R}^m} e^{-\frac{1}{2\hbar}p\cdot M p+\frac{i}{\hbar}v\cdot p}\,dp=(\det\, M)^{-1/2}\exp\Big(-\frac{1}{2\hbar}v\cdot M^{-1}v\Big) \ . 
\end{equation}
In the above we take $\sqrt{\cdot }$ to be the principal branch of the square root function.

We now consider the asymptotic behavior of oscillating integrals of the form 
\begin{equation}
\Big(\frac{i}{2\pi\hbar}\Big)^{m/2}\int_{\mathbb{R}^m} a(p)e^{\frac{i}{\hbar}S(x,p)}\,dp 
\end{equation}
for small $\hbar>0$ and $x\in \mathbb{R}^n$.
\\

\noindent \textbf{Theorem} (Nazaikinksii, Oshmyan, Sterin and Shatalov \cite{NOSS})\textbf{.} -- \textit{Consider the semi-classical Fourier oscillatory integral for $\hbar>0$,} 
\begin{equation}
\Big(\frac{i}{2\pi\hbar}\Big)^{m/2}\int a(p)e^{\frac{i}{\hbar}S(x,p)}\,dp\end{equation}
\textit{where $a:\mathbb{R}^{m}\rightarrow \mathbb{C}$ and $S:\mathbb{R}^n\times \mathbb{R}^m \rightarrow \mathbb{C}$ are of class $C^s$ in $p$, where $a$ is compactly supported and $S$ possesses an everywhere non-negative imaginary part on ${\rm supp}\,a$, ${\rm Im}\,S\geq 0$, so that the equations}
\begin{equation}
{\rm Im}\,S(x,p)=0 \ , \ \ \frac{\partial S}{\partial p}(x,p)=0
\end{equation}
\textit{have at most a single solution on ${\rm supp}\,a$ for given $x\in\mathbb{R}^n$, denoted by $p=p(x)\in\mathbb{R}^m$, while the Hessian matrix}
\begin{equation}
S''(x,p):=\frac{\partial^2S}{\partial p^2}(x,p)
\end{equation}
\textit{is non-singular for all $(x,p)\in\mathbb{R}^n\times \mathbb{R}^m$. Then, for all $r=0,\ldots, s$, the following semi-classical relation holds as $\hbar\rightarrow 0^+$}
\begin{equation}
\fl \Big(\frac{i}{2\pi\hbar}\Big)^{m/2}\int a(p)e^{\frac{i}{\hbar}S(x,p)}\,dp=e^{\frac{i}{\hbar}\,{}^r\!S(x,z (x))}\frac{{}^r\!a(z (x))}{\sqrt{{\rm det}\,-\,{}^r\!S''(x,z (x))}}\Bigg(1+o(\hbar)\Bigg)
\end{equation}
\textit{where $\sqrt{\cdot }$ is the principal branch of the square root function, $\,{}^r\!a(z)$ and $\,{}^r\!S(x,z)$ the $r$-analytic extensions of $a$ and $S$, respectively, to the complex variable $z:=p+i\xi$ and $\xi\in\mathbb{R}^m$, ${}^r\!S(x,z )$ and $\,{}^r\!S''(x,z ):=\frac{\partial^2\,{}^r\!S}{\partial z ^2}(x,z )$ are the $r$-analytic extensions of $S$ and $S''$ to the complex variable $z =p+i\xi$, and $z =z (x)$ for $x\in\mathbb{R}^n$ is the unique complex solution of the equation} 
\begin{equation}
\frac{\partial S}{\partial z }(x,z )=0 \ . 
\end{equation}
\textit{In addition, there exists $c>0$ such that} 
\begin{equation}
{\rm Im}\, S(x,z(x))\geq c\,|{\rm Im}\,z(x)|^2
\end{equation}
\textit{for all $x\in\mathbb{R}^n$.} $\Box$
\\

Finally, we state the commutation formula for phase space semi-classical Weyl operators with complex phases, stating, first, the basic result for semi-classical Weyl operators with real phases. 
\\

\noindent \textbf{Theorem} (Karasev and Maslov \cite{KaMa})\textbf{.} -- \textit{Consider a semi-classical Weyl operator $\widehat f$ with Weyl symbol $f\in C^\infty(\mathbb{R}^{2n}, \mathbb{R})$ and a phase $S\in C^\infty(\mathbb{R}^{n},\mathbb{R})$. Then, as $\hbar\rightarrow 0^+$, we have that}
\begin{eqnarray}
\fl e^{-\frac{i}{\hbar}S}f\Big(\stackrel{\boldsymbol{\omega}}{x},-i\hbar\, \stackrel{\boldsymbol{\omega}}{\frac{\partial}{\partial x}}\Big) e^{\frac{i}{\hbar}S}(x=f\Big(x,\frac{\partial S}{\partial x}(x)\Big)\\ \nonumber -\frac{i\hbar}{2} \,{\rm tr}\,\Bigg(\frac{\partial^2 f}{\partial x\partial p}\Big(x,\frac{\partial S}{\partial x}(x)\Big)+\frac{\partial^2 f}{\partial p^2}\Big(x,\frac{\partial S}{\partial x}(x)\Big)\,\frac{\partial^2 S}{\partial x^2}(x)\Bigg)+O(\hbar^2) \ . \ \ \Box
\end{eqnarray}

We generalize to semi-classical Weyl operators on double phase space, with symbol 
\begin{equation}
\sigma_W(\widecheck H)(X,P)=H\Big(\frac{X}{2}-JP\Big)
\end{equation}
based on the Commutation Theorem of Kucherenko \cite{Kuc1}. 
\\

\noindent \textbf{Theorem.} -- \textit{Consider a physical quantity $H\in C^\infty(\mathbb{R}^{2d},\mathbb{R})$ and a phase $F={\rm Re}\,F+i\,{\rm Im}\,F\in C^\infty(\mathbb{R}^{2d},\mathbb{C})$. Then, for all $s\in \mathbb{N}$, as $\hbar\rightarrow 0^+$, we have that} 
\begin{eqnarray}
\fl H\Big(\stackrel{\boldsymbol{\omega}}{\frac{X}{2}}+i\hbar\, J\stackrel{\boldsymbol{\omega}}{\frac{\partial}{\partial X}}\Big)e^{\frac{i}{\hbar}F(X)} =e^{\frac{i}{\hbar}F(X)}\Bigg(\,{}^r\! H\Big(\frac{X}{2}-J\frac{\partial F}{\partial X}(X)\Big)\\ \nonumber
+\frac{i\hbar}{2}\, {\rm tr}\,\frac{\partial^2 F}{\partial X^2} \,{}^{s }\!\Big(\frac{\partial^2 H}{\partial X^2}\Big)\Big(\frac{X}{2}-J\frac{\partial F}{\partial X}(X)\Big)\Bigg)+r(X;\hbar)
\end{eqnarray}
\textit{where} 
\begin{equation}
\fl \,{}^r\! H\Big(\frac{X}{2}-J\frac{\partial F}{\partial X}(X)\Big)=\sum_{\alpha\in\mathbb{N}_0^{2d}}\sum_{|\alpha|=0}^r\frac{1}{\alpha!}\Big(-iJ\frac{\partial {\rm Im}\,F}{\partial X}(X)\Big)^\alpha(\partial_X^\alpha H)\Big(\frac{X}{2}-J\frac{\partial {\rm Re}\,F}{\partial X}(X)\Big)
\end{equation}
\textit{while the remainder $r(X;\hbar)$ satisfies the growth condition given in the Commutation Theorem of Kucherenko.} $\Box$

\bigskip

{\bf Bibliography}
\\

\end{document}